\documentclass[a4paper,fleqn,usenatbib, useAMS]{mnras}
\usepackage{graphicx}
\usepackage{multicol}
\usepackage{enumerate}
\usepackage{amsmath}
\usepackage{pdflscape}
\usepackage{amssymb}
\usepackage{longtable}
\usepackage[flushleft]{threeparttable}

\usepackage{journal_names}
\def\farcs{\hbox{$.\!\!^{\prime\prime}$}}
\def\kms{km s$^{-1}$}
\title[BH masses of optically/UV selected TDEs] {Black hole masses of tidal disruption event host galaxies}
\author[Wevers et al.]{Thomas Wevers$^{1}$\thanks{Email: t.wevers@astro.ru.nl}, Sjoert van Velzen$^{2}$, Peter G. Jonker$^{3,1}$, Nicholas C. Stone$^{4}$, \newauthor Tiara Hung$^{5}$, Francesca Onori$^{3}$, Suvi Gezari$^{5}$ and Nadejda Blagorodnova$^{6}$ \\\\
$^{1}$Department of Astrophysics/IMAPP, Radboud University, P.O. Box 9010, 6500GL Nijmegen, The Netherlands\\
$^{2}$Department of Physics \& Astronomy, The Johns Hopkins University, Baltimore, MD 21218, USA\\
$^{3}$SRON, Netherlands Institute for Space Research, Sorbonnelaan 2, 3584CA Utrecht, The Netherlands\\
$^{4}$Columbia Astrophysics Laboratory, Columbia University, New York, NY, 10027, USA\\
$^{5}$Department of Astronomy, University of Maryland, College Park, MD 20742, USA\\
$^{6}$Cahill Center for Astrophysics, California Institute of Technology, Pasadena, CA 91125, USA\\
}

\begin{document}
\date{}
\pagerange{\pageref{firstpage}--\pageref{lastpage}} \pubyear{2017}
\maketitle
\label{firstpage}
\begin{abstract}
The mass of the central black hole in a galaxy that hosted a tidal disruption event (TDE) is an important parameter in understanding its energetics and dynamics. We present the first homogeneously measured black hole masses of a complete sample of 12 optically/UV selected TDE host galaxies (down to $g_{host}$\,$\leq$\,22 mag and $z$\,=\,0.37) in the Northern sky. The mass estimates are based on velocity dispersion measurements, performed on late time optical spectroscopic observations. We find black hole masses in the range 3\,$\times$\,10$^5$ M$_{\odot}$\,$\leq$\,M$_{\rm BH}$\,$\leq$\,2\,$\times$\,10$^7$ M$_{\odot}$. The TDE host galaxy sample is dominated by low mass black holes ($\sim$\,10$^6$ M$_{\odot}$), as expected from theoretical predictions. The blackbody peak luminosity of TDEs with M$_{\rm BH}$\,$\leq$\,10$^{7.1}$ M$_{\odot}$ is consistent with the Eddington limit of the SMBH, whereas the two TDEs with M$_{\rm BH}$\,$\geq$\,10$^{7.1}$ M$_{\odot}$ have peak luminosities below their SMBH Eddington luminosity, in line with the theoretical expectation that the fallback rate for M$_{\rm BH}$\,$\geq$\,10$^{7.1}$ M$_{\odot}$ is sub-Eddington. In addition, our observations suggest that TDEs around lower mass black holes evolve faster. These findings corroborate the standard TDE picture in 10$^6$ M$_{\odot}$ black holes. Our results imply an increased tension between observational and theoretical TDE rates. By comparing the blackbody emission radius with theoretical predictions, we conclude that the optical/UV emission is produced in a region consistent with the stream self-intersection radius of shallow encounters, ruling out a compact accretion disk as the direct origin of the blackbody radiation at peak brightness. 
\end{abstract}

\section{Introduction}
\label{sec:introduction}
It is currently accepted that supermassive black holes (SMBH) reside in the centers of most, if not all, massive galaxies (e.g. \citeauthor{Kormendy1995} \citeyear{Kormendy1995}). If there is gas close to the hole, its accretion has directly observable signatures and we designate the center an active galactic nucleus (AGN). However, if there is no gas near the SMBH, indirect methods must be used to infer its presence. Occasionally a reservoir of gas may wander near the black hole in the form of a star. If the tidal forces due to the SMBH are larger than the self-gravity of the star, the SMBH will tear it apart, and about half of the star will be accreted by the central black hole \citep{Rees1988, Phinney1989, Evans1989}. This so-called tidal disruption of a star is accompanied by a luminous flare at X-ray, UV or optical wavelengths, announcing the presence of an otherwise dormant SMBH to the Universe. 

In the last two decades, about two dozen tidal disruption events (TDEs) have been discovered in various wavelength regimes such as X-rays \citep{Donley2002, Komossa2002, Cenko2012, Maksym2013}, UV \citep{Gezari2008, Gezari2009} and optical light \citep{vanVelzen2011, Gezari2012, Arcavi2014, Chornock2014, Holoien2016}. From an observational point of view, there seem to be two broad classes of TDEs: those where X-ray (or even higher energy) emission was detected, and those where optical emission was detected. It should be noted that not all optical TDEs were followed up at X-ray wavelengths, which may partially explain this apparent dichotomy. Two exceptions are already known, including ASASSN--15oi \citep{Holoien15oi} and ASASSN--14li, which was detected not only at optical \citep{Holoien2016} and X-ray \citep{Milleretal2015} wavelengths but was also observed to produce radio emission \citep{Alexander2016, vanVelzen2016}. 

In the classical picture of TDEs, the electromagnetic radiation is produced when the bound debris circularizes and falls back to the SMBH \citep{Rees1988, Phinney1989}. An accretion disk forms at a radius of about 2\,R$_{\rm p}$, where R$_{\rm p}$ is the pericenter radius of the orbit of the disrupted star. The disk forms rapidly and efficiently circularizes due to stream-stream collisions induced by relativistic precession. While this scenario is able to explain the properties of TDEs producing X-rays, the temperatures and luminosities of optical TDEs are an order of magnitude lower than theoretical predictions \citep{vanVelzen2011}. Several scenarios have been proposed to explain the optical emission mechanism of TDEs, including thermal reprocessing of accretion power by material far from the hole \citep{Loeb1997, Guillochon2014}, shock emission produced by the self-intersecting debris stream \citep{Piran2015} or outflows \citep{Strubbe2009, Miller2015, Metzger2016, Stone2016}. More recently, magnetic stresses have also been considered as the source of both X-ray and optical emission \citep{Bonnerot2017}. A theoretical framework that can explain the dynamics and energetics of both X-ray and optical emission from TDEs has yet to converge towards a unified theory.

Observational studies of TDEs are critical to provide meaningful constraints on key ingredients for theoretical models, such as the dynamical efficiency of stream circularization, the primary TDE power source, and the dominant emission mechanisms. Because of the two-body nature of a TDE, constraining the mass of the black hole component helps to disentangle other aspects of the events, including the dynamics and energetics. For instance, the tidal radius of the disrupted star, the energetics of the accretion phase, the post-disruption dynamics and the expected electromagnetic (and gravitational wave) emission all depend on the black hole mass. Constraining the black hole mass can also provide direct constraints on the accretion efficiency or the amount of mass accreted during a TDE. Currently the mass of the black hole is usually inferred from modelling rather than used as an input parameter because no accurate, systematic measurements are available. 

Constraining the mass of a black hole in the center of a galaxy has a rich history (see \citeauthor{Ferrarese2005} \citeyear{Ferrarese2005} for a review). The discovery of correlations between the bulge luminosity and mass (the M\,--\,L relation, e.g. \citeauthor{Dressler1989} \citeyear{Dressler1989}, \citeauthor{Kormendy1995} \citeyear{Kormendy1995}) or bulge velocity dispersion and mass (the M\,--\,$\sigma$ relation, e.g. \citeauthor{Ferrarese2000} \citeyear{Ferrarese2000} or \citeauthor{Gebhardt2000} \citeyear{Gebhardt2000}) indicate that there is a tight connection between the evolution and formation of the SMBH and the stellar bulge \citep{Kormendy2013}. 
By exploiting these correlations, it is possible to measure black hole masses even when it is not possible to spatially resolve the sphere of influence of the SMBH (at $z$\,$\geq$\,0.01) and derive the mass from the dynamics of stars or gas that is directly influenced by the black hole. At higher redshifts, using these scaling relations has the advantage of being less data intensive than direct methods such as reverberation mapping. They have therefore made SMBH mass measurements a relatively easy task (compared to direct methods) at redshifts in excess of $z$\,$\sim$\,0.01.

A robust method for extracting the velocity dispersion from galaxy spectra is to compare the width and equivalent width of stellar absorption lines with stellar template libraries in pixel space (e.g. \citeauthor{Rix1992} \citeyear{Rix1992}, \citeauthor{vdM1994} \citeyear{vdM1994}, \citeauthor{Cappellari2004} \citeyear{Cappellari2004}). Working in pixel space makes masking bad pixels more easy, while it also facilitates the simultaneous modelling of gas and stellar kinematics with other observational effects such as contamination due to emission-line gas \citep{Cappellari2017}.

In this work we present the first systematic effort to measure the black hole masses of a sample of 12 optically/UV selected TDE host galaxies. In Section \ref{sec:observations}, we describe the sample selection and observations used to perform the measurements. Section \ref{sec:analysis} explains the methodology we followed; we present the results and discuss their implications in Section \ref{sec:results}. Finally, we summarize in Section \ref{sec:summary}.

\section{Observations and data reduction}
\label{sec:observations}
\subsection{Host galaxy sample}
We have obtained spectroscopic observations (Table \ref{tab:observations}) of galaxies hosting optically/UV selected {\it nuclear} transients with a blackbody temperature in excess of 10$^4$\,K (which we will refer to as TDEs) located in the Northern sky (declination $\geq$ 0$^{\circ}$). Our sample is complete down to a limiting (host galaxy) magnitude of $g_{\rm host}$\,=\,22 mag; the hosts span a range in redshift from 0.016 to 0.37. These transients were discovered by a variety of surveys (see Table \ref{tab:observations} for references to the discovery papers), including the Sloan Digital Sky Survey (SDSS), the All Sky Automated Survey for Supernova (ASAS--SN), the (intermediate) Palomar Transient Factory (PTF), the Panoramic Survey Telescope and Rapid Response System (PS1) and the Galaxy Evolution Explorer (GALEX). Our sample comprises 12 sources out of a total of 13 optically/UV discovered TDEs in the Northern sky\footnote{http://TDE.space}. PS1--11af is the remaining source at $g_{host}$\,=\,23 and z\,=\,0.405 \citep{Chornock2014}. There is one TDE in our sample for which a discovery article has not yet been published in the literature: iPTF--15af. This TDE was discovered in the galaxy SDSS J084828.13+220333.4 (\citeauthor{French2016} \citeyear{French2016}). \\

The observations were performed with the William Herschel Telescope (WHT, Section \ref{sec:wht}) on La Palma, Spain, the Very Large Telescope (VLT, Section \ref{sec:vlt}) at Cerro Paranal, Chile and the Keck--II telescope on Mauna Kea, Hawaii. 
\begin{table*}
 \centering
  \caption{Overview of the observations used in this work. The galaxies are sorted according to increasing redshift. Slit gives the slit width used, and $\sigma_{\rm instr}$ is the instrumental broadening (in km s$^{-1}$) as measured from sky or arc lamp lines. The value of $\sigma_{\rm instr}$ is calculated at 3900\,\AA\ in the rest-frame of the host except for D3--13, where it is given at 5000\,\AA\ (because of the rest-frame wavelength coverage of the spectrum).}
  \begin{tabular}{cccccccc}
  \hline
  Name & RA & Decl. & Telescope & Instrument & Slit &$\sigma_{instr}$ & Reference \\
  \hline
iPTF--16fnl &00:29:57.01 &32:53:37.2 & VLT & X-shooter/UVB &  0\farcs8 & 20 & \citet{Blago2017}  \\
ASASSN--14li &12:48:15.23 & 17:46:26.4 & WHT & ISIS/R600 & 0\farcs8 & 50 &  \citet{Holoien2016} \\
&&&Keck & ESI &  0\farcs5 & 16 & \\
ASASSN--14ae & 11:08:40.12&34:05:52.2 & WHT & ISIS/R600 &  0\farcs7 & 40 & \citet{Holoien2014} \\
&&&Keck & ESI &  0\farcs5 & 16 & \\
PTF--09ge &14:57:03.18 &49:36:41.0 & WHT & ISIS/R600 &  1\farcs1 & 55 & \citet{Arcavi2014} \\
&&&Keck & ESI &  0\farcs5 & 16 & \\
iPTF--15af & 08:48:28.13 & 22:03:33.4 & Keck & ESI & 0\farcs5 & 16 & \citet{French2016}\\
iPTF--16axa & 17:03:34.34 & 30:35:36.7 & Keck & ESI & 0\farcs5 & 16 & \citet{Hung2017}\\
PTF--09axc & 14:53:13.08&22:14:32.3  & WHT & ISIS/R600 & 1\farcs1 & 55 & \citet{Arcavi2014} \\
SDSS TDE1 & 23:42:01.41&01:06:29.3 &  WHT & ISIS/R600 &  1\farcs1 & 55 &  \citet{vanVelzen2011} \\
PS1-10jh&16:09:28.28 & 53:40:24.0 & Keck & ESI & 0\farcs5 & 16 & \citet{Gezari2012}\\
PTF--09djl& 16:33:55.97 & 30:14:16.6  & WHT & ISIS/R600 &  1\farcs1 & 55 &  \citet{Arcavi2014} \\ 
&&&Keck & ESI &  0\farcs5 & 16 & \\
GALEX D23H--1 &23:31:59.54& 00:17:14.6& WHT & ISIS/R600 &  1\farcs1  & 55 & \citet{Gezari2009} \\
GALEX D3--13 &14:19:29.81 & 52:52:06.4& Keck & DEIMOS/1200G & 1\farcs0 & 35 & \citet{Gezari2006} \\
\hline
    \end{tabular}
  \label{tab:observations}
\end{table*}

\subsection{WHT/ISIS}
\label{sec:wht}
We obtained late time spectra of some TDE host galaxies using the Intermediate dispersion Spectrograph and Imaging System (ISIS, \citeauthor{Jorden1990} \citeyear{Jorden1990}) mounted at the Cassegrain focus of the 4.2m William Herschel Telescope (WHT) located on La Palma, Spain. We used the R600B and R600R gratings in the blue and red arm respectively, with central wavelengths optimized for covering wavelength regions containing host galaxy absorption lines. There is a gap in the coverage between the blue and red arms due to the use of a dichroic at 5300\,\AA. The wavelength coverage of this setup is 1000\,\AA\ around the central wavelength of each arm. A summary of the observations is presented in Table \ref{tab:observations}. 

\begin{figure*} 
  \includegraphics[height=9cm, keepaspectratio]{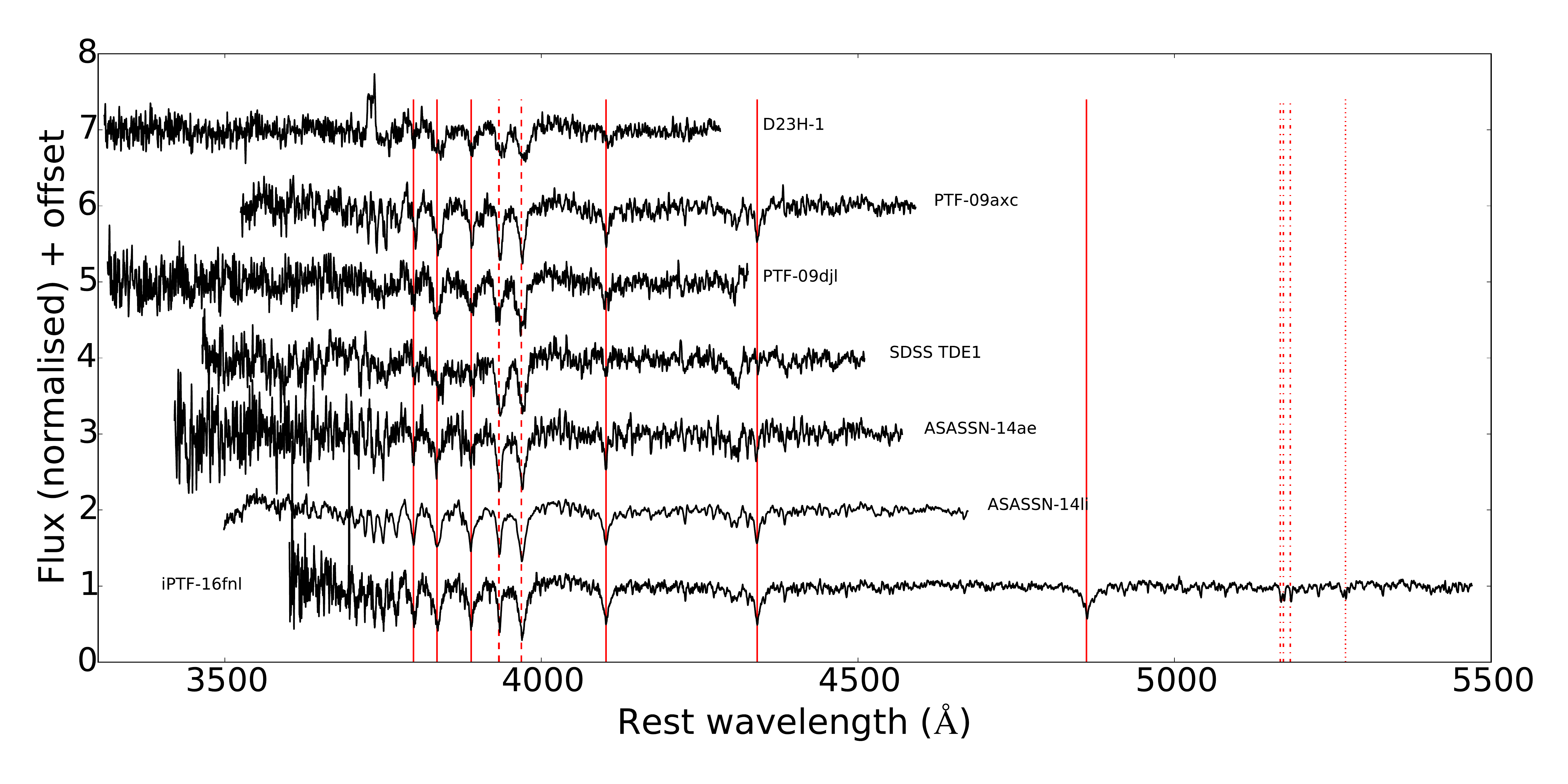}
  \includegraphics[height=9cm, keepaspectratio]{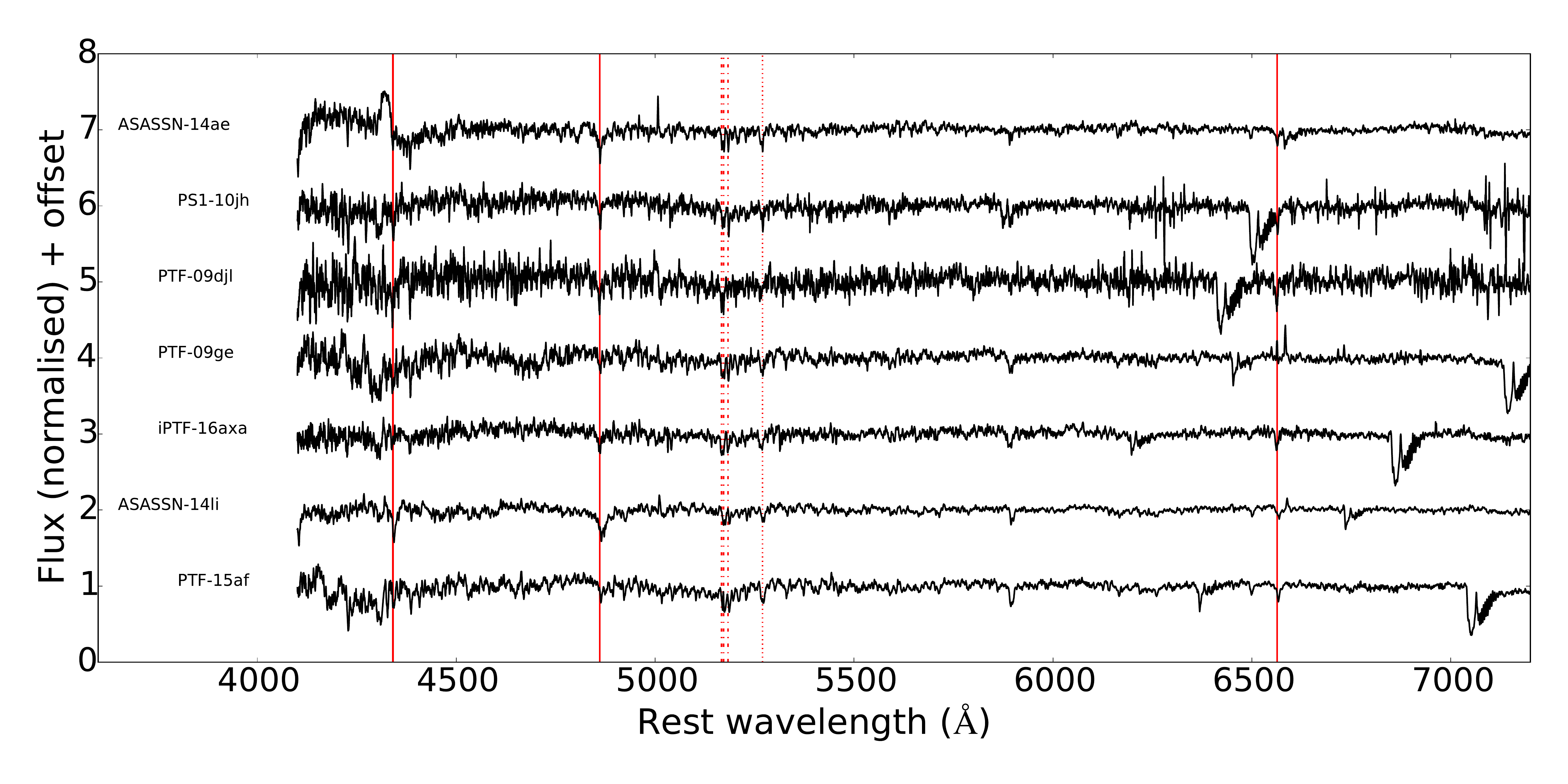}
    \caption{Top panel: continuum normalized TDE host galaxy spectra. The top 6 spectra were taken with WHT/ISIS (blue arm), while the bottom spectrum was taken with VLT/X-shooter (UVB arm). The spectra are shifted to the rest-frame wavelength of the hosts. Solid lines mark transitions of the H Balmer series. The two dashed lines mark the Ca H and K lines at $\lambda\lambda$3934,3968. The dash-dotted and dotted lines mark the Mg\,\textsc{i}b and Fe $\lambda$5270 lines. Bottom: same, but showing the Keck/ESI spectra. The spectra have been smoothed with a boxcar filter with a 10 pixel width for display purposes. The noise in the red part of the PS1--10jh and PTF--09djl spectra is due to incomplete sky line subtractions. We only show the part of the spectrum that was used for template fitting.}
 \label{fig:observations}
\end{figure*}

We first perform the standard reduction steps such as a bias level subtraction, a flat field correction and a wavelength calibration using \textsc{iraf}. Cosmic rays are removed using the \textit{lacos} package in \textsc{iraf} \citep{vanDokkum2012}. The typical root-mean-square deviation (rms) of the applied wavelength solution is $\leq$\,0.1 \AA, which corresponds to at most 0.5 pixels. The absolute wavelength calibration is evaluated by measuring the position of a Hg\,\textsc{i} sky line at $\lambda4358.33$, and when necessary the spectra are shifted to match the same wavelength scale. This ensures that combining multiple spectra of the same source does not introduce an artificial broadening of the absorption lines. The spectra are rebinned to a linear dispersion on a logarithmic wavelength scale. We perform an optimal extraction \citep{Horne1986}, which weights each pixel along the spatial profile by the inverse variance of the number of detected photons (i.e. pixels containing less signal get down-weighted) to achieve the highest possible signal-to-noise ratio (SNR) for the extracted spectrum. The variance spectra are also calculated and will be used for Monte Carlo simulations (Section \ref{sec:analysis}). 
We measure the instrumental broadening of the different observational setups using arc lamp observations taken together with the science spectra to measure $\sigma_{\rm instr}$. The resolution of the observations is slit-limited for all spectra. Our observations provide an instrumental resolution FWHM of 1.75 \AA\ in the blue arm for a 1\farcs1 slit width (or better, if the slit width was smaller), which corresponds to 55 km s$^{-1}$ at 3900\,\AA\ (Table \ref{tab:observations}). We present the resulting spectra in Figure \ref{fig:observations} (top panel). 

\subsection{VLT/X-shooter}
\label{sec:vlt}
For iPTF--16fnl, we have obtained a late time spectrum ($\sim$\,193 days after peak brightness) in which the TDE does not contribute a significant fraction to the total galaxy light on 2016 November 25 (Onori et al. in prep.) with X-shooter \citep{Vernet2011}, mounted on UT2 (Kueyen) of the Very Large Telescope (VLT) at Cerro Paranal, Chile. The 1800\,s observation (OB ID: 1617353) was performed using an 0\farcs8 slit. The spectral resolution provided by this setup is R\,=\,6200, which yields an instrumental broadening equivalent to $\sigma$\,=\,20 km s$^{-1}$ at 3900\,\AA. We use the ESO Phase 3 pipeline\footnote{http://www.eso.org/observing/dfo/quality/XSHOOTER/pipeline} reduced spectrum of the UVB arm for our analysis, which has an absolute wavelength calibration accurate to 0.3\,\AA.

\subsection{Keck/ESI}
\label{sec:keck}
We took medium resolution spectra with the Echelette Spectrograph and Imager (ESI; \citeauthor{Sheinis2002} \citeyear{Sheinis2002}), mounted at the Cassegrain focus of the Keck--II telescope on Mauna Kea, Hawaii. The instrument provides a wavelength coverage ranging from 3900\,--\,10000\,\AA\ in multiple echelle orders. The observations were performed using a 0\farcs5 slit, providing a near-constant resolving power of R\,=\,8000. The FWHM resolution is 38 km s$^{-1}$, which translates to an instrumental resolution of $\sigma_{\rm instr}$\,=\,16 km s$^{-1}$.

The data were reduced using the MAuna Kea Echelle Extraction (\texttt{makee}) software package, which was developed and optimised for the reduction of ESI data. The pipeline performs standard spectroscopic data reduction routines including bias subtraction, flatfielding and spectrum extraction. The standard star Feige 34 was used to compute the trace of the science objects. The position of each echelle order is traced, optimally extracted and wavelength calibrated independently, after which the different orders are rebinned to a linear dispersion on a logarithmic wavelength scale with a dispersion of 11.5 km s$^{-1}$ per pixel. The orders are combined using the {\it combine} command to produce a 1D spectrum. The wavelength calibration is performed in \textsc{iraf} using two arc lamp (CuAr and HgNe+Xe) exposures.

\subsection{Further data processing}
After obtaining the 1D spectra from our WHT, VLT and Keck observations, further processing steps are required before we can measure the velocity dispersion. The spectra are normalized by fitting 3$^{\rm rd}$ order cubic splines to the continuum in \textsc{molly}. We mask all prominent absorption and emission lines during this process to identify the continuum. We average the spectra, weighting by the mean SNR (variance) of each individual exposure. 

We extract spectra from two different spatial regions of the host galaxy for each exposure (see Section \ref{sec:comparisonofsigma}). One extraction includes the whole galaxy along the slit, to increase the SNR of the resulting spectrum. The second extraction region is centered on the peak of the light profile, and has an aperture radius equal to the seeing of the exposure. This extraction aims at isolating as much as possible the bulge region of the galaxy, to provide an estimate of the central velocity dispersion rather than the luminosity-weighted velocity dispersion obtained from the entire galaxy. We measure the seeing using point sources present on the slit; if not available, we use measurements of a local seeing monitor (the Robotic Differential Image Motion Monitor, available for the WHT data) as an estimate. In case no measurements are available, we use an aperture equal to the slit width, effectively mimicking a square fiber with sides equal to the slit width.

\section{Velocity dispersion measurements}
\label{sec:analysis}
We use the penalized pixel fitting (\textsc{ppxf}) method \citep{Cappellari2004, Cappellari2017} to measure the line of sight velocity dispersion function (LOSVD), typically denoted as $f(v)$, of the galaxies in our sample. Briefly, the method consists of convolving a set of template spectra with an initial guess for $f(v)$, which is then compared to the observed host galaxy spectrum. The LOSVD is parametrized by a series of Gauss-Hermite polynomials in the form:
\begin{equation}
f(v) = \frac{1}{\sigma \sqrt{2\pi}}\ \text{exp}\left(\frac{1}{2}\frac{(v-V)^2}{\sigma^2}\right) \bigg[ 1 + \sum_{m=3}^{M} h_m\ H_m(\frac{v-V}{\sigma}) \bigg]
\end{equation}
where V is the mean velocity along the line of sight, $\sigma$ is the velocity dispersion, H$_m$ are Hermite polynomials and h$_m$ their coefficients. The Hermite polynomials are defined as
\begin{equation}
H_i = \frac{1}{\sqrt{i!}} e^{x^{2}} \Big(-\frac{1}{\sqrt{2}}\frac{\partial}{\partial x} \Big) e^{-x^{2}}
\end{equation}
\label{eq:losvd}
where we include terms up to H$_4$. The terms H$_3$ and H$_4$ parametrize the asymmetric and symmetric deviations from a Gaussian line profile, respectively. 
The best-fitting template is found by $\chi^2$ minimization, using the set of templates convolved with $f(v)$ for the variables [V, $\sigma$, h$_3$, h$_4$]. The \textsc{ppxf} method was specifically designed to extract accurate kinematical information in the case of low SNR spectra. We refer the reader to \citet{Cappellari2004} and \citet{Cappellari2017} for more details. 

\subsection{Template library}
We note that the red part of the WHT spectra does not contain well defined, deep and unblended absorption lines suitable for a robust measurement of the velocity dispersion. At bluer wavelengths, the Ca\,\textsc{ii} H+K absorption lines at $\lambda\lambda3934,3968$ in combination with many smaller absorption lines provide the best means to determine the velocity dispersion. The H Balmer absorption lines are known to be strongly affected by pressure broadening due to collisional or ionizational excitation, and we exclude them from the measurement process. We therefore only use the blue part of the WHT spectra, starting at 3900\,\AA. We fit the full spectral range, as the use of many absorption lines present in the spectrum will improve the measurement of the velocity dispersion. We mask the H Balmer lines, and in addition emission lines of O\,\textsc{iii} at $\lambda\lambda4959,5007$, the diffuse interstellar band at $\lambda5780$ and the interstellar Na\,\textsc{i}\,D absorption lines at $\lambda\lambda5890,5895$. 

Based on the highest resolution spectrum and the wavelength coverage of the observations, we choose template spectra from the ELODIE v3.1 database \citep{Prugniel2001, Prugniel2007}. This spectral library contains 1554 templates at R\,=\,10000 at 5500\,\AA, which implies a velocity dispersion resolution of $\sigma$\,=\,17 \kms\ at 3900\,\AA. By using a large set of templates we minimize the effects of mismatches between the observed galaxy spectra and the templates used to derive the line broadening. The best-fitting parameters are obtained by $\chi^2$ minimization. Because the higher order terms (h$_3$ and h$_4$) can only be robustly constrained in the case of high SNR data, the method includes a bias factor which penalizes these terms in the best-fitting solution to 0 in case the SNR is low. We follow the procedure outlined in \citet{Emsellem2004} to determine the appropriate value for the penalty in the fitting procedure for each galaxy. 

During the measurement process (in \textsc{ppxf}) for the Keck spectra, we take into account that the template FWHM resolution (in \AA) is independent of wavelength (0.54 \AA), but the ESI spectral resolution (in \AA) varies with wavelength. We only use the wavelength range where $\sigma_{\rm template}$\,$\leq$\,$\sigma_{\rm ESI}$, starting at 4300 \AA\ and ending at 6800 \AA, where the template spectral coverage stops. 

\subsection{Luminosity-weighted LOSVD and central LOSVD}
\label{sec:comparisonofsigma}
In contrast with the IFU/fiber observations that are typically used to measure the kinematics of galaxies (for example in the SDSS Baryonic Oscillations Spectroscopic Survey, \citealt{Dawson2013}), we measure the LOSVD using long-slit observations. For spectroscopic observations obtained using a fiber instrument with a $\sim$\,few arcsec diameter, one expects an evolution of the measured velocity dispersion with the ability to spatially resolve the bulge of the galaxy, i.e. with redshift (e.g. \citeauthor{Bernardi2003} \citeyear{Bernardi2003}). For increasing distances, the velocity dispersion is influenced by stars at larger physical radii, and thus depends on the velocity dispersion profile of the galaxy. We use longslit observations, and the measurements including the entire galaxy in the extraction region are effectively luminosity-weighted velocity dispersions. It was shown by \citet{Gebhardt2000} that such measurements reflect the central velocity dispersion to good degree (to within 5 per cent, see their figure 1) as long as the slit width is smaller than or comparable to the effective light radius of the host galaxy.

It should be noted that the sample used by \citet{Gebhardt2000} consists of galaxies at much lower redshifts and with higher masses. Therefore the bulge region in these nearby, massive elliptical galaxies is more dominant in a long-slit observation than we expect them to be for our sample, which consists of galaxies at higher redshifts and smaller bulge masses, as theory predicts these smaller SMBHs to produce higher rates of TDEs \citep{Magorrian1999, Wang2004, Stone2016}. The underlying principle still holds, but the luminosity-weighted LOSVD measurements of our sample must be interpreted with care: its relation to the central velocity dispersion depends on the relative dominance of the bulge region over the rest of the galaxy. For this reason, we provide central velocity dispersion measurements based on the careful extractions outlined in Section \ref{sec:observations}, which aim at isolating the velocity dispersion in the central part of the galaxy. 

\subsection{Robust velocity dispersions}
To robustly estimate the velocity dispersion and its uncertainty induced by the measurements, we perform 1000 Monte Carlo simulations. We resample the original spectrum by drawing flux values from a Gaussian distribution within the errors as obtained from the optimal extraction for each pixel. This ensures that the data quality of each simulation (i.e. the average SNR) remains the same and does not influence our measurements. We fit the resulting distribution of velocity dispersion values with a Gaussian function, and adopt the mean and standard deviation as the best-fitting value for $\sigma$ and its uncertainty.
\section{Results and discussion}
\label{sec:results}
\begin{figure*} 
  \includegraphics[width=0.8\textwidth,keepaspectratio]{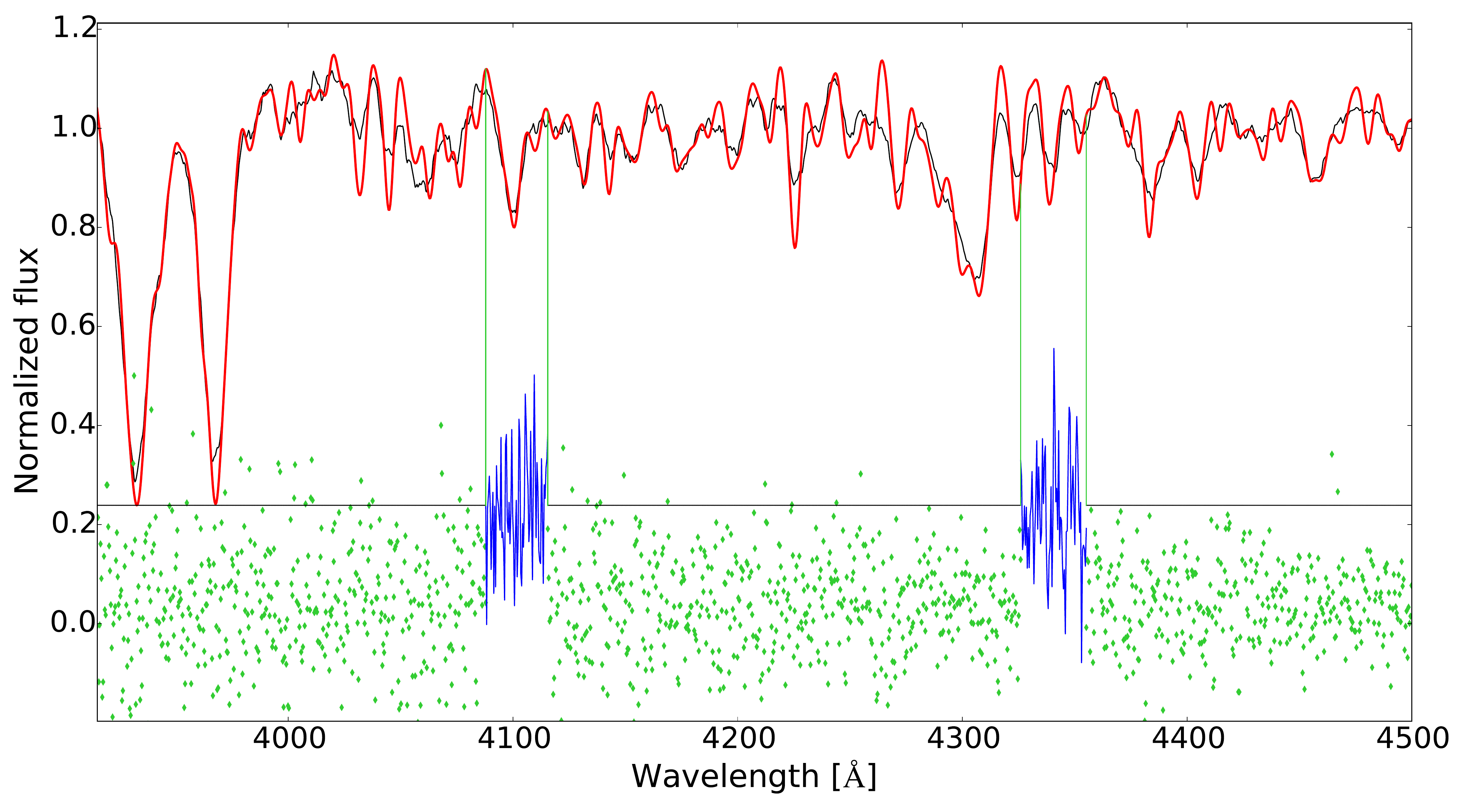}
  \caption{Part of the continuum normalized WHT spectrum of TDE1, overlaid with the best-fitting template spectrum (red) broadened to a velocity dispersion of 126 \kms. The residuals are shown in green. Blue regions are excluded from the fit.}
  \label{fig:14li}
\end{figure*}

\begin{figure} 
  \includegraphics[width=0.5\textwidth, keepaspectratio]{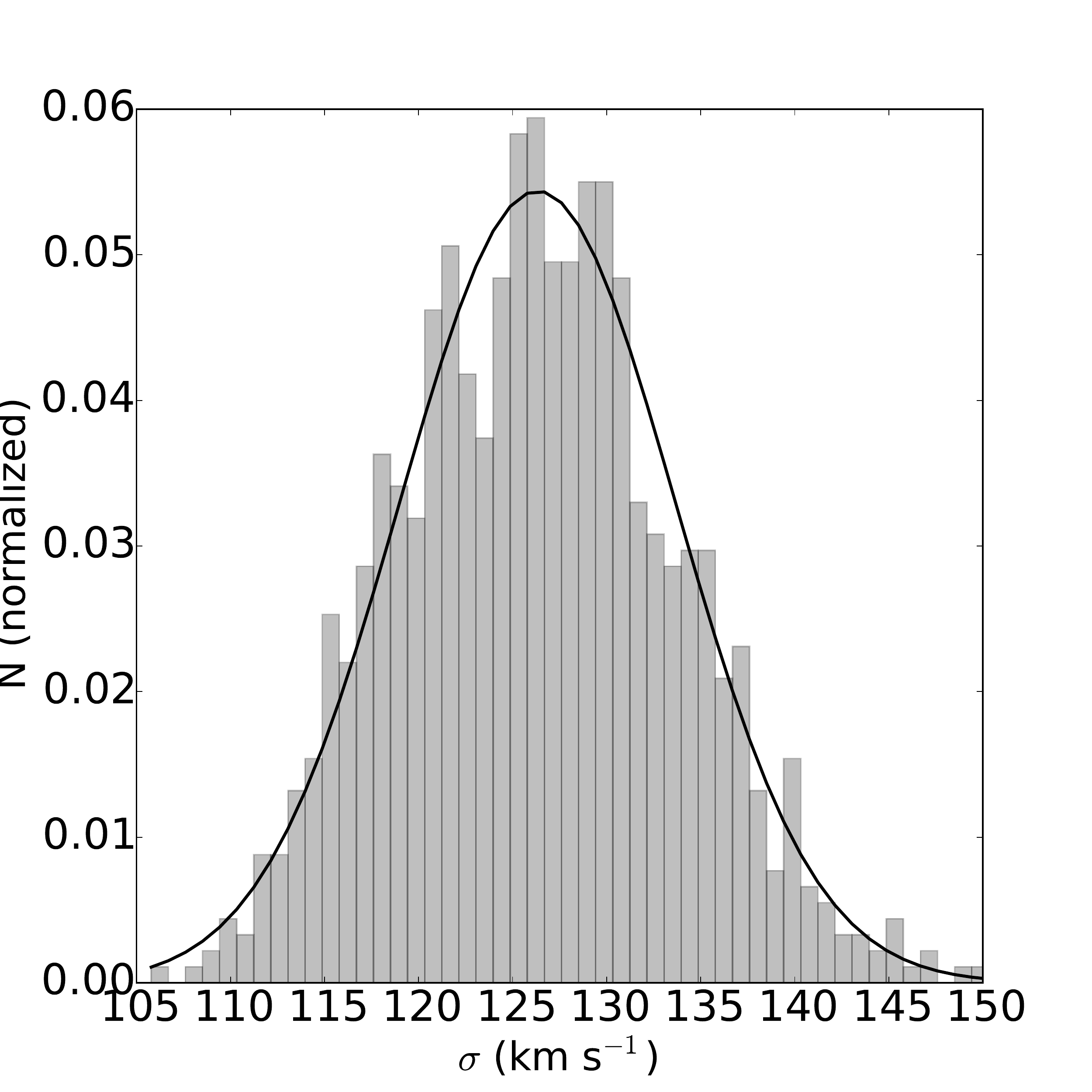}
  \caption{Distribution of velocity dispersion measurements for TDE1 obtained from 1000 Monte Carlo trials of the WHT spectrum. The distribution is well approximated by a Gaussian, with a mean value of 126 \kms\ and a standard deviation of 7 \kms.}
  \label{fig:14lihistogram}
\end{figure}

As an illustration, we show the result of the template fitting procedure in Figure \ref{fig:14li} using the WHT spectrum of TDE1. Overlaid in red is the best-fitting template spectrum broadened to 126 km s$^{-1}$. The residuals are shown in green, while blue regions are excluded in the fitting process. The velocity dispersion is well defined and the fit describes the data well, leaving little structure in the residuals. In Figure \ref{fig:14lihistogram} we show the distribution of measured $\sigma$ values and the Gaussian fit used to determine the mean and standard deviation. 

To obtain black hole masses, we assume that the M\,--\,$\sigma$ relation holds for all the velocity dispersions we measure, and convert the measurements to masses using the relation from \citet{Ferrarese2005}:
\begin{equation}
\label{eq:msigma}
\centering
\frac{M_{\rm BH}}{10^8 M_{\odot}} = 1.66 \times \left(\frac{\sigma}{200\ \text{km s}^{-1}}\right)^{4.86}
\end{equation}
To estimate the uncertainties in the black hole mass, we add the uncertainties of the velocity dispersion measurements linearly with the 0.34 dex systematic uncertainty introduced by using the M\,--\,$\sigma$ relation \citep{Ferrarese2005}. The uncertainty is dominated by the scatter in the M\,--\,$\sigma$ relation except for D23H--1.
In Table \ref{tab:sigmas} we present the results of the velocity dispersion measurements for our sample. We also include the redshift, host galaxy magnitude and half-light radius, as well as literature values of velocity dispersion measurements for comparison purposes. 

\begin{table*}
 \centering
  \caption{Measured central velocity dispersions ($\sigma_{\rm WHT/VLT}$ and $\sigma_{\rm Keck}$) and nuclear black hole masses using the relation from \citet{Ferrarese2005}. When the central velocity dispersion could not be determined, we use the luminosity-weighted value to calculate the black hole mass and mark the value with a *. For PTF--09djl, we deem $\sigma_{\rm WHT}$ unreliable (see text). The uncertainty on the mass is the linear sum of the systematic uncertainty from the M\,--\,$\sigma$ relation (0.34 dex) and the error introduced by the measurement uncertainty. We also include the host $g$-band (Petrosian) magnitude and the bulge half-light radius (from a deVaucouleur profile fit) in the $g$-band from SDSS. Slit gives the slit width, in arcseconds. $\sigma_{\rm lit}$ are literature values, taken from SDSS \citep{Thomas2013} except for D3--13 and iPTF--16fnl, where we quote the values by \citet{Gezari2006} and \citet{Blago2017}, respectively. We omit SDSS measurements below 70 km s$^{-1}$ as they are unreliable.}
  \begin{tabular}{ccccccccc}
  \hline
  Name & $\sigma_{\rm WHT/VLT}$ & $\sigma_{\rm Keck}$ &  log$_{10}$(M$_{\rm BH}$) & m$_g$ &Half-light radius & Slit & $\sigma_{\rm lit}$\\
  & km s$^{-1}$ & km s$^{-1}$ &M$_{\odot}$ & & arcsec &  arcsec & \kms \\
  \hline
\vspace{1mm}iPTF--16fnl & 55\,$\pm$\,2  & --- & 5.50$^{+0.42}_{-0.42}$ & 15.61 & 5.2 & 0.8& 89\,$\pm$\,1 \\\vspace{1mm}
ASASSN--14li & 72\,$\pm$\,3 & 81\,$\pm$\,2 & 6.23$^{+0.39}_{-0.40}$ &  16.15& 1.0 & 0.8/0.5& --- \\\vspace{1mm}
ASASSN--14ae &  56\,$\pm$\,7 & 53\,$\pm$\,2 & 5.42$^{+0.46}_{-0.46}$ & 17.49 & 2.9& 0.7/0.5&--- \\\vspace{1mm}
PTF--09ge & 72\,$\pm$\,6 & 82\,$\pm$\,2  & 6.31$^{+0.39}_{-0.39}$ & 18.06 & 2.8& 1.1/0.5& ---  \\\vspace{1mm}
iPTF--15af & --- & 106\,$\pm$\,2 & 6.88$^{+0.38}_{-0.38}$ & 18.64 & 1.9 & 0.5 &98\,$\pm$\,11 \\\vspace{1mm}
iPTF--16axa & --- & 82\,$\pm$\,3 & 6.34$^{+0.42}_{-0.42}$ &  19.46 & 1.7 & 0.5&---\\\vspace{1mm}
PTF--09axc  & 60\,$\pm$\,4 & --- & 5.68$^{+0.48}_{-0.49}$ & 18.87 & 0.5 & 1.1&--- \\\vspace{1mm}
SDSS TDE1 & 126\,$\pm$\,7 & --- & 7.25$^{+0.45}_{-0.46}$ & 20.44& 0.6 & 1.1&137\,$\pm$\,12 \\\vspace{1mm}
PS1--10jh & --- & 65\,$\pm$\,3 & 5.85$^{+0.44}_{-0.44}$ &  21.95 & 0.26 &0.5& ---\\\vspace{1mm}
PTF--09djl & 104\,$\pm$\,13 & 64\,$\pm$\,7 & 5.82$^{+0.56}_{-0.58}$ & 20.72 & 0.3 & 1.1/0.5&--- \\ \vspace{1mm}
GALEX D23H--1 & 77\,$\pm$\,18$^{*}$ & --- & 6.21$^{+0.78}_{-0.90}$  & 20.23 & 0.6 & 1.1& 86\,$\pm$\,14\\
GALEX D3--13 & 133\,$\pm$\,6$^{*}$ & --- & 7.36$^{+0.43}_{-0.44}$ &21.99& 0.6 &1.0& 120\,$\pm$\,10 \\
\hline
    \end{tabular}
  \label{tab:sigmas}
\end{table*}
\subsection{Comparison to independent measurements}
\label{sec:independent}
For several sources in our sample, velocity dispersion measurements are available in the literature. In Table \ref{tab:sigmas} we list the literature values alongside our own measurements. Several of the velocity dispersions measured from SDSS spectra are below the instrumental resolution, which we deem less reliable, especially for low SNR observations. Three sources can be reliably compared: TDE1, D23H--1 and iPTF--15af. We quote the measurements performed by \citet{Thomas2013} as these authors also use \textsc{ppxf} to measure $\sigma$, although they use a different set of templates and a different wavelength regime (4500\,--\,6500\,\AA). For TDE1 these authors find $\sigma$\,=\,137\,$\pm$\,12 \kms, while we find a slightly smaller value of $\sigma$\,=\,126\,$\pm$\,7 \kms. The measured values for D23H--1 and iPTF--15af are consistent within the errors with the SDSS measurements of \citet{Thomas2013}. The velocity dispersion of D3--13 was measured using a similar template fitting procedure by \citet{Gezari2006}, and was measured to be 120\,$\pm$\,10 km s$^{-1}$. Using our resampling approach we find $\sigma$\,=\,133\,$\pm$\,6 \kms, slightly higher but consistent within the mutual uncertainties. We also note that for iPTF--16fnl there is a discrepancy between our measured value (55\,$\pm$\,2 km s$^{-1}$) and that of \citet{Blago2017} (89\,$\pm$\,1 km s$^{-1}$), who fit Gaussian lines to the Mg\,\textsc{i}\,b and Ca\,\textsc{ii} triplet simultaneously.  

Furthermore, we have WHT and Keck spectra of 4 sources, providing another opportunity for independent measurements. For ASASSN-14ae we measure 56\,$\pm$\,7 and 53\,$\pm$\,2 km s$^{-1}$ using the ISIS and ESI spectra, respectively, while for ASASSN--14li we measure 72\,$\pm$\,3 and 81\,$\pm$\,2 km s$^{-1}$. We use the inverse-variance weighted average of these independent measurements as the best estimate of the velocity dispersion: $\sigma_{\rm avg}$\,=\,53\,$\pm$\,2 \kms\ and $\sigma_{avg}$\,=\,78\,$\pm$\,2 \kms\ for ASASSN--14ae and ASASSN--14li, respectively. For PTF--09ge, we calculate an inverse-variance weighted mean of $\sigma_{avg}$\,=\,81\,$\pm$\,2 \kms.
Regarding PTF--09djl, there appears to be an inconsistency of $\sim$\,40 km s$^{-1}$ between the Keck (64\,$\pm$\,7 km s$^{-1}$) and WHT (104\,$\pm$\,13 km s$^{-1}$) values. We note that the overlapping wavelength coverage of the WHT spectrum with the templates is small ($\sim$\,500 \AA), and a visual inspection of the best-fitting template with the galaxy spectrum reveals that the fit is poor. Moreover, our WHT spectra use a 1\farcs1 arcsec slit width, while the bulge half-light radius of this galaxy is 0\farcs3 and hence does not satisfy the criterion of \citet{Gebhardt2000} (see discussion below). On the other hand, the best-fitting solution to the Keck spectrum is satisfactory. We therefore adopt the value as measured from the Keck spectrum as the best representation of the central velocity dispersion of this source. 

\subsection{Potential caveats}
\subsubsection{Signal-to-noise ratio (SNR) and $\sigma$}
We have determined the value and uncertainty of $\sigma$ by performing Monte Carlo simulations (Table \ref{tab:sigmas}). We find that, as expected, the accuracy with which $\sigma$ can be recovered is strongly dependent on the SNR and the wavelength coverage of the data. 
\begin{figure} 
  \includegraphics[width=0.475\textwidth, keepaspectratio]{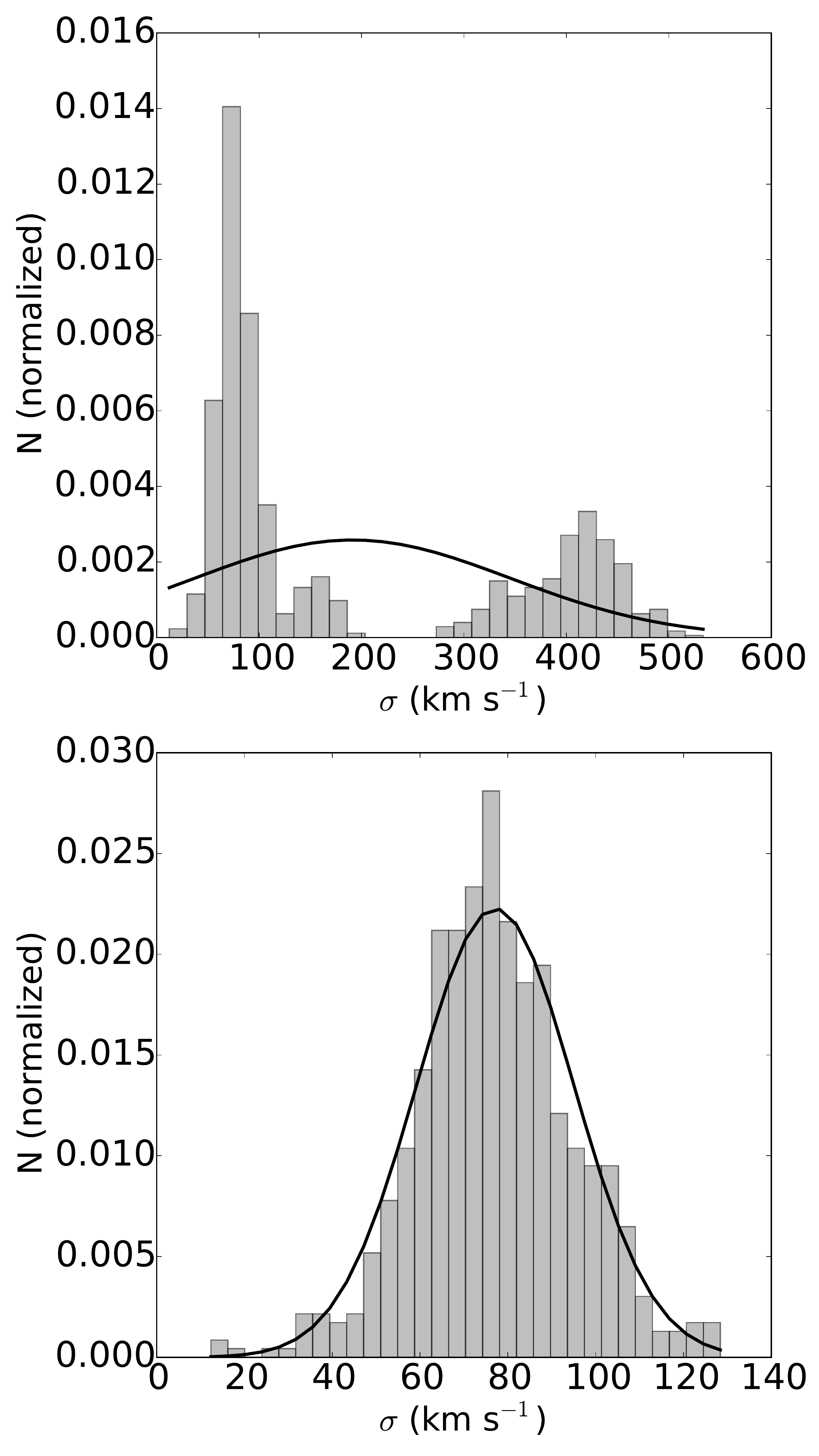}
  \caption{Distribution of the Monte Carlo trials to determine the best-fitting velocity dispersion of D23H--1. Top: including all trials; bottom: including trials limited to $\sigma$\,=\,[0,130] km s$^{-1}$ (see text). }
  \label{fig:mctrialsd23h1}
\end{figure} 
For the spectrum of D23H--1, the relatively low SNR of the spectra causes a degeneracy in the best-fitting velocity dispersion. Due to the large errors in the observed spectrum, the $\chi^2$ minimization is not able to resolve the shallow, narrow absorption lines. Instead, the minimization procedure finds a good fit with larger values of $\sigma$\,$\sim$\,few hundred \kms, essentially fitting only a few broad absorption lines instead of the myriad of low SNR, low equivalent width absorption lines present in the spectrum. In this case, we use the trials corresponding to a limited (but conservatively large) range of $\sigma$ to determine the best-fitting velocity dispersion. We perform Monte Carlo simulations until this limited range contains at least 1000 trials, to robustly estimate the uncertainty induced by the measurement errors.
 An illustration is shown in Figure \ref{fig:mctrialsd23h1} for the WHT spectrum of D23H--1, including a fit to all the trials (top) and a fit to only the trials in the range $\sigma$\,=\,[0,130] km s$^{-1}$ (bottom). We adopt $\sigma$\,=\,77\,$\pm$\,18 \kms\ in this case. We also note that extracting the central region of the host galaxy to D23H--1 results in a low SNR spectrum. The model fitting becomes less constrained and we are no longer able to robustly measure the central velocity dispersion of this galaxy.

For the Keck spectra, the large wavelength coverage makes it is possible to accurately determine the velocity dispersion even with a relatively low SNR per pixel because of the large number of small lines in the spectrum. The large wavelength coverage (hence large number of degrees of freedom), combined with the fact that no very deep absorption lines are present (such as the Ca H+K lines in the WHT spectra) also makes the template selection procedure more prone to make non-optimal choices. This can lead to a divergence in the fitting process if there are only a few absorption lines in common between the science spectrum and the selected templates to determine $\sigma$. This is observed as a tail of outliers at high velocity dispersion values; we therefore use a similar procedure as outlined above for D23H--1 to fit a Gaussian to a restricted range in the velocity dispersion distribution. 

\subsubsection{Comparison of luminosity-weighted and central LOSVDs}
We find no significant differences between the luminosity-weighted LOSVDs and the central velocity dispersion values. In all cases the measurments yield results that are consistent within the mutual errors. In Table \ref{tab:sigmas} we provide the host galaxy half-light radius, as determined by SDSS \citep{Stoughton2002} from a de Vaucouleur profile fit to the galaxy light. We note that for all sources except for the WHT spectrum of PTF--09djl, our observations are within the regime where the slit width is less than two times the galaxy half-light radius, for which \citet{Gebhardt2000} have shown that the luminosity-weighted LOSVD is a good tracer of the central velocity dispersion. For the WHT spectra of PTF--09djl we are not in this regime (as discussed in Section \ref{sec:independent}). We therefore adopt the value obtained from the Keck spectrum, obtained with a slit width of 0\farcs5, as the most reliable measurement. 

For the other sources, we do not find significant differences between the luminosity-weighted and central LOSVDs, implying as expected that our long-slit data, even when extracting the full galaxy light, are not strongly influenced by the disk of the galaxy. We note that using an optimal extraction for the spectra will have helped in this respect.

\subsubsection{Choice of M\,--\,$\sigma$ relation}
\label{sec:msigma}
\begin{figure} 
  \includegraphics[width=0.475\textwidth, keepaspectratio]{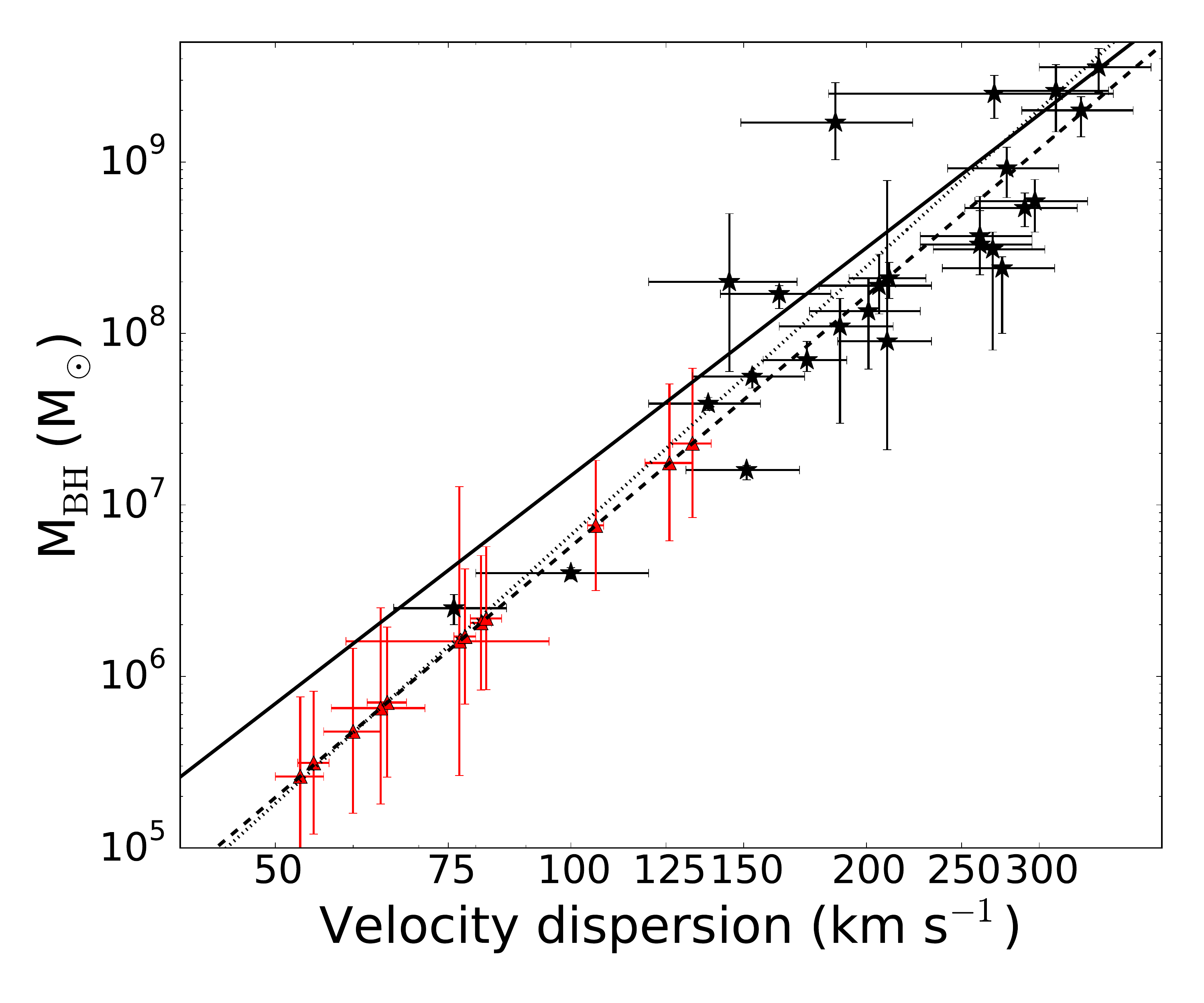}
  \caption{TDE host black hole masses and various versions of the M\,--\,$\sigma$ relation. Black stars represents the resolved sample of \citet{Ferrarese2005}, while the dashed line represents the best-fitting relation (Eq. \ref{eq:msigma}); red triangles represent the TDE host galaxies. The dotted line represents the \citet{Mcconnell2013} relation valid for early type galaxies. The solid line represents the \citet{Kormendy2013} relation for massive ellipticals. Regarding the latter relation, we remark that our galaxies are not ellipticals and therefore it is unlikely that this relation is appropriate for our sample.}
  \label{fig:msigma_overplotted}
\end{figure}

The particular choice of M\,--\,$\sigma$ relation and which version is the {\it best} version is still a matter of debate, with many versions published in the literature \citep{Ferrarese2005, Gultekin2009, Mcconnell2013, Kormendy2013}. Each of these works has its particular sample selection that comes with advantages and disadvantages. In this work we have chosen to use the relation based on the sample of \citet{Ferrarese2005}, who included only galaxies for which the sphere of influence had been resolved. If we compare these values with those obtained with the recent \citet{Mcconnell2013} relation, valid for early-type galaxies, we find that the (non-systematic) difference is less than 0.1 dex for the sources in our sample. Therefore we do not expect the particular choice of the M\,--\,$\sigma$ relation to influence our conclusions. In Figure \ref{fig:msigma_overplotted} we show the original (resolved) sample used by \citet{Ferrarese2005} to derive the M\,--\,$\sigma$ relation (Eq. \ref{eq:msigma}; dashed line). We have overplotted the relation by \citet{Mcconnell2013} (dotted line) and \citet{Kormendy2013} (solid line) to illustrate the effect on the derived masses. We note that the latter relation was explicitly derived for elliptical galaxies and is most likely not appropriate for our sample.

Another issue that arises from using the M\,--\,$\sigma$ relation for our sample is that several host galaxies harbour black holes with masses that are lower than the mass range for which the relation was originally derived (see also Figure \ref{fig:msigma_overplotted}). Simulations have shown that the (currently unknown) black hole seed formation scenario has an impact on the validity of the  M\,--\,$\sigma$ relation at the low mass end. For example, \citet{Volonteri2010} showed that in the case of high-mass seeds the relation should show an increased scatter, possibly combined with a flattening at low $\sigma$. However, there is at present no conclusive evidence that corroborates these predictions. For example, \citet{Barth2005} measure black hole masses for less than 10$^6$ M$_{\odot}$ BHs and find that they lie on the extrapolation of the M\,--\,$\sigma$ relation to lower masses. \citet{Xiao2011} found that the relation derived for quiescent massive ellipticals can also be extrapolated to active galaxies, with masses as low as 2\,$\times$\,10$^5$\,M$_{\odot}$. These authors did not find evidence for an increased scatter in the correlation at the low end of the mass range. We remark that direct mass measurements for these systems are needed to resolve this issue beyond doubt.

\begin{figure} 
  \includegraphics[width=0.5\textwidth, keepaspectratio]{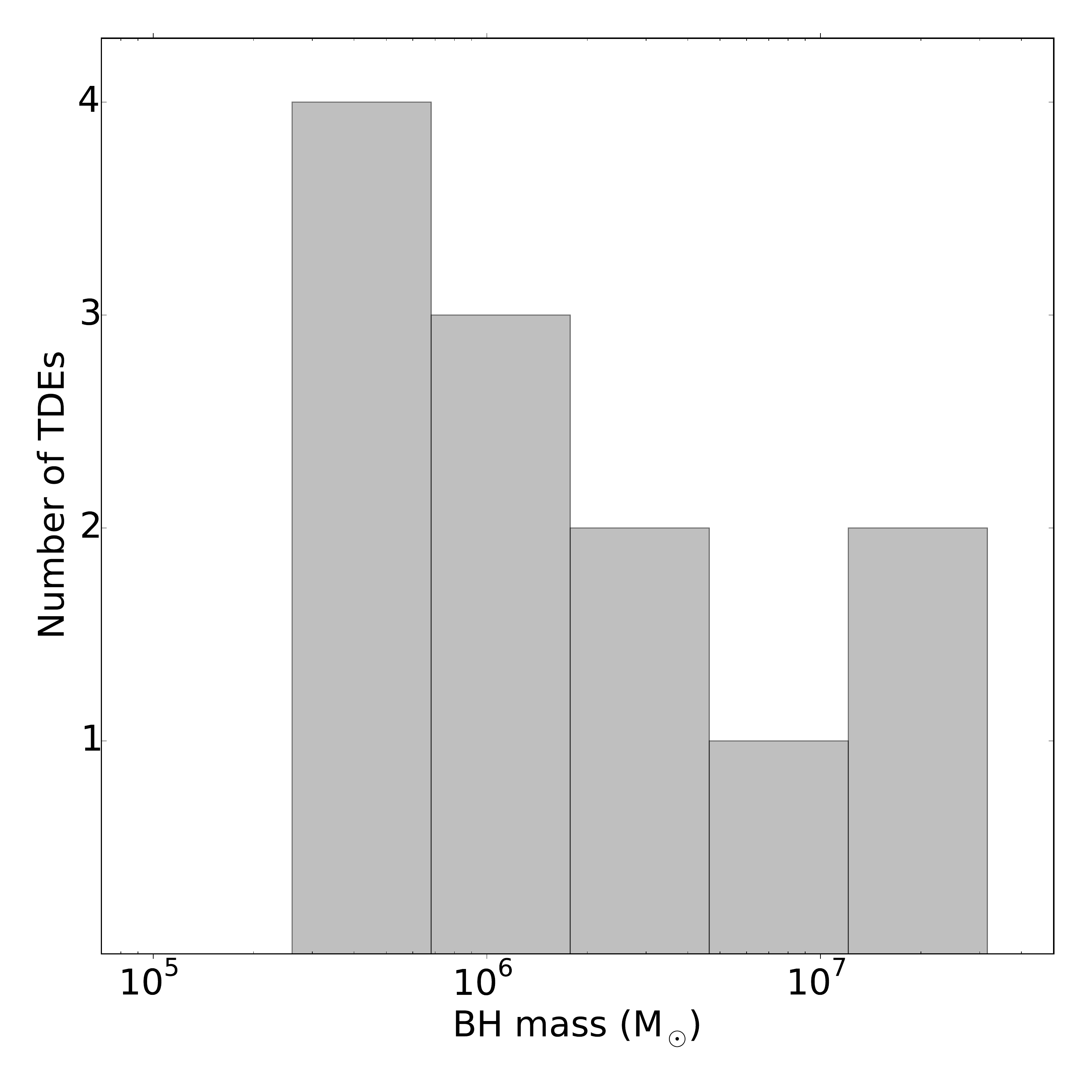}
  \caption{Distribution of the observed black hole masses in our sample of TDE host galaxies. The sample is dominated by low mass black holes, as expected from theoretical arguments \citep{Wang2004}. This is in contrast to fig. 12 of \citet{Stone2016}, who found a more top-heavy $M_{\rm BH}$ distribution, with SMBH masses in optical TDE hosts peaked just below $10^7 M_{\odot}$.}
  \label{fig:massdistrib}
\end{figure} 
\subsection{A black hole mass distribution for TDE host galaxies}
\label{sec:massfunction}
Recent theoretical work has used the observed sample of TDE candidates to analyze flare demographics \citep{Kochanek2016}, to constrain the SMBH occupation fraction in low mass galaxies \citep{Stone2016}, and to try to constrain optical emission mechanisms \citep{Stone2016, Metzger2016}. The BH/bulge mass estimates used in these works are inhomogeneous, but are generally based on the M\,--\,L relation, and the bulge mass of these galaxies is subject to large uncertainties. Here we present a new and updated black hole mass distribution based on spectroscopic measurements of our host galaxy sample. 

Our mass distribution, presented in Figure \ref{fig:massdistrib}, contains black hole masses ranging from 3\,$\times$\,10$^{5}$ M$_{\odot}$ to 2\,$\times$\,10$^{7}$ M$_{\odot}$. It is dominated by low mass black holes in the range $\sim$\,10$^6$ M$_{\odot}$. The absence of black holes with masses lower than 3\,$\times$\,10$^{5}$ M$_{\odot}$ could be explained by the increasingly smaller volume in which TDEs can still be detected around low mass black holes (assuming that the peak luminosity is Eddington-limited or otherwise scales with the black hole mass). Alternatively, this could be a consequence of the black hole occupation fraction in low mass galaxies \citep{Stone2016} or because of a lower flare luminosity due to inefficient circularization \citep{Dai2015, Guillochon2015}. On the high mass end, the lack of SMBHs in excess of 10$^{7.5}$ M$_{\odot}$ could be explained by the direct capture of stars \citep{Hills1975}. Testing this hypothesis requires a careful treatment of the survey completeness due to both the host and TDE flux limits, and will be explored in detail in van Velzen et al. (in prep.).

\begin{table*}
 \centering
  \caption{Host galaxy and TDE properties of our sample. We have included the velocity dispersion and derived black hole mass, host redshift, Eddington luminosity, integrated blackbody luminosity, blackbody temperature, decay rate and (K-corrected) peak absolute magnitude in the $g$-band. All logarithms are with base 10. Values between brackets indicate the uncertainty in the last digit. The uncertainties in the Eddington luminosity are identical to the uncertainties in the black hole mass and are omitted from the table. Values marked with a $^*$ are lower limits. We also give the reference work from which data were taken. For iPTF--15af no data are available in the literature.}
   \begin{threeparttable}  
  \begin{tabular}{ccccccccccc}
  \hline\vspace{1mm}
  Name & $\sigma$  &  log(M$_{\rm BH}$)  & $z$ & log(L$_{\rm Edd}$) & log(L$_{\rm BB}$) & T$_{\rm BB}$ & R$_{\rm BB}$ & Decay rate & M$_g$ & Ref.  \\
  & km s$^{-1}$ & M$_{\odot}$ & & erg s$^{-1}$ & erg s$^{-1}$ & 10$^3$\,K & 10$^{14}$\,cm &mag/100d  & mag & \\
  \hline\vspace{1mm}
  iPTF--16fnl & 55\,$\pm$\,2  & 5.50$^{+0.42}_{-0.42}$ & 0.016 & 43.6 & 43.5(1)&35(3.5) & 1.8(4)  & 4.4\,$\pm$\,0.3 & --17.2 & a,b \\\vspace{1mm}
ASASSN--14li & 78\,$\pm$\,2 & 6.23$^{+0.39}_{-0.40}$ & 0.021 & 44.3 & 43.8(1) &35(3) & 2.4(5) & 0.92\,$\pm$\,0.05 & --17.7 & c,d \\\vspace{1mm}
ASASSN--14ae &  53\,$\pm$\,2 & 5.42$^{+0.46}_{-0.46}$ & 0.043 &43.5 & 43.9(1) &21(2) & 7(1.5) &  1.7\,$\pm$\,0.3 & --19.1 & e \\\vspace{1mm}
PTF--09ge &  81\,$\pm$\,2  & 6.31$^{+0.39}_{-0.39}$ & 0.064 & 44.4 & 44.1(1) & 22(2) & 9(2) & 1.58\,$\pm$\,0.04& --19.9 & f  \\\vspace{1mm}
iPTF--15af & 106\,$\pm$\,2 & 6.88$^{+0.38}_{-0.38}$ &0.079 & 45.0 & --- & --- & --- & ---&---&--- \\\vspace{1mm}
iPTF--16axa & 82\,$\pm$\,3 & 6.34$^{+0.42}_{-0.42}$ & 0.108 &44.4 & 44.5(1) &30(3) & 7.6(1.5) & 1.85\,$\pm$\,0.07 & --19.1 &g \\\vspace{1mm}
PTF--09axc  & 60\,$\pm$\,4  & 5.68$^{+0.48}_{-0.49}$ & 0.115& 43.8 & 43.49(5) &12(1) & 14.5(3) & 0.7$^*$ & --19.5 & f  \\\vspace{1mm}
SDSS TDE1 & 126\,$\pm$\,7 & 7.25$^{+0.45}_{-0.46}$ & 0.136 & 45.4 &43.5(1)&24(3) & 3.6(1) & 1.7\,$\pm$\,0.3 & --18.1 & h \\\vspace{1mm}
PS1--10jh & 65\,$\pm$\,3 & 5.85$^{+0.44}_{-0.44}$ & 0.170 &44.0 & 44.21(7) & 29(2) & 5.7(9) & 2.56\,$\pm$\,0.07 & --19.4 & i\\\vspace{1mm}
PTF--09djl & 64\,$\pm$\,7  & 5.82$^{+0.56}_{-0.58}$ & 0.184& 43.9 & 44.4(1) &26(3) & 9(2) & 0.6$^*$  & --20.2 & f \\\vspace{1mm}
GALEX D23H--1 & 77\,$\pm$\,18 &  6.21$^{+0.78}_{-0.90}$ & 0.185& 44.3 & 44.0(1) & 49(5) & 1.5(4) &0.67\,$\pm$\,0.04 & --17.3 & j \\
GALEX D3--13 & 133\,$\pm$\,6 & 7.36$^{+0.43}_{-0.44}$ & 0.369 & 45.5 & 44.30(5) &49(2) & 2.2(2) & 0.26\,$\pm$\,0.02 & --18.2 & j,k \\
\hline
    \end{tabular}
      \begin{tablenotes}
  \small \item $^{a}$\citet{Blago2017}, $^{b}$ \citet{Brown2017}, $^{c}$\citet{Holoien2016}, $^{d}$\citet{vanVelzen2016}, $^{e}$\citet{Holoien2014}, $^{f}$\citet{Arcavi2014}, $^{g}$\citet{Hung2017}, $^{h}$\citet{vanVelzen2011}, $^{i}$\citet{Gezari2012}, $^{j}$\citet{Gezari2009}, $^{k}$\citet{Gezari2006}
  \end{tablenotes}
 \end{threeparttable}
  \label{tab:correlations}
\end{table*}

We remark that our mass distribution is in contrast with masses taken from the literature (e.g. figure 12 of \citealt{Stone2016}). These authors found a more top-heavy $M_{\rm BH}$ distribution peaked just below $10^7 M_{\odot}$, with SMBH masses mostly derived using the M\,--\,L relation. We list a few potential explanations for this difference below. To start, \citet{Stone2016} did not apply B/T corrections for most galaxies, implying that the resulting masses are upper limits. A second potential caveat is that many TDE host galaxies are rare E+A galaxies \citep{Arcavi2014, French2016}, which are thought to possess a central overdensity of stars due to a recent merger \citep{Zabludoff1996}. These galaxies are observed to have very centrally peaked light profiles (see e.g. \citealt{vanVelzenStone2016}), and therefore they could be overluminous with respect to the galaxies used to derive the scaling relation (typically massive ellipticals). This was also noted by \citet{French2017} as a caveat to their analysis, and may explain why we find lower BH masses for 3 sources (ASASSN--14ae, ASASSN--14li and PTF--09ge) with M$_{\rm BH}$ estimated from M$_{\rm bulge}$ using stellar population fitting (\citealt{French2017}; their table 2). Finally, \citet{Graham2012} have shown that the M\,--\,L relation may be a broken power law rather than applicable to the whole mass range; they find that it should have a steeper slope (M\,$\propto$\,L$^{2}$ instead of M\,$\propto$\,L$^{1}$) below $\sim$\,10$^{8}$ M$_{\odot}$. This would lead to an overestimate of $M_{\rm BH}$ for masses below $\sim$\,10$^{8}$ M$_{\odot}$. Based on numerical simulations, \citet{Fontanot2015} identified that stellar feedback due to star formation may lead to a change of slope in the M\,--\,L scaling relation. \citet{Graham2015} also suggest that a steeper relation can explain the presence of samples of low mass AGNs with seemingly undermassive BHs.

\subsection{Correlations with other observables}
Recent studies investigating potential correlations between the black hole mass and other TDE observables such as peak luminosity and $e$-folding timescale are reported by \citet{Hung2017} and \citet{Blago2017} respectively. Despite some suggestive evidence, no strong correlations were observed. However, this could be a consequence of the heterogeneous mass measurements available in the literature, motivating us to re-investigate potential correlations. In Figure \ref{fig:correlations} we plot our black hole masses against other observables. We provide the plotted data in Table \ref{tab:correlations}. We search for correlations between the observables using the Spearman rank correlation metric. Similar to previous work, we do not find statistically significant (95 per cent confidence interval) correlations. This could be a consequence of the small sample size, in combination with the degeneracy of different parameters such as the mass of the star and the impact parameter. Nevertheless, it is instructive to discuss some suggestive evidence for correlations with the host black hole mass or derived Eddington luminosity. It is important to note that our galaxy sample is drawn from flux-limited surveys, and we do not consider the effects of a flux limit for the flare itself. We will find that the qualitative trends corroborate the tidal disruption interpretation of these events, and moreover can provide input and constraints for viable TDE emission models.

\subsubsection{Redshift}
Figure \ref{fig:correlations} (panel a) suggests that TDEs found at lower redshift are associated with lower mass black holes. The dearth of TDEs found in low mass black holes at higher redshifts may be a consequence of the flux limited nature of our sample. The lack of higher black hole masses for TDE hosts at low redshifts could be explained by the relative rarity of higher mass black holes, as the log(N)\,--\,log(M) distribution of black hole masses rises towards lower masses (e.g. \citeauthor{Shankar2009} \citeyear{Shankar2009}). The exponential tail of the black hole mass function implies that a large volume is needed to include enough high mass black holes. As a result, in a flux limited sample the observed black hole mass distribution is expected to correlate with redshift as long as it does not contain a representative sample of galaxies.

\subsubsection{Peak absolute magnitude}
\begin{figure*} 
  \includegraphics[width=\textwidth, keepaspectratio]{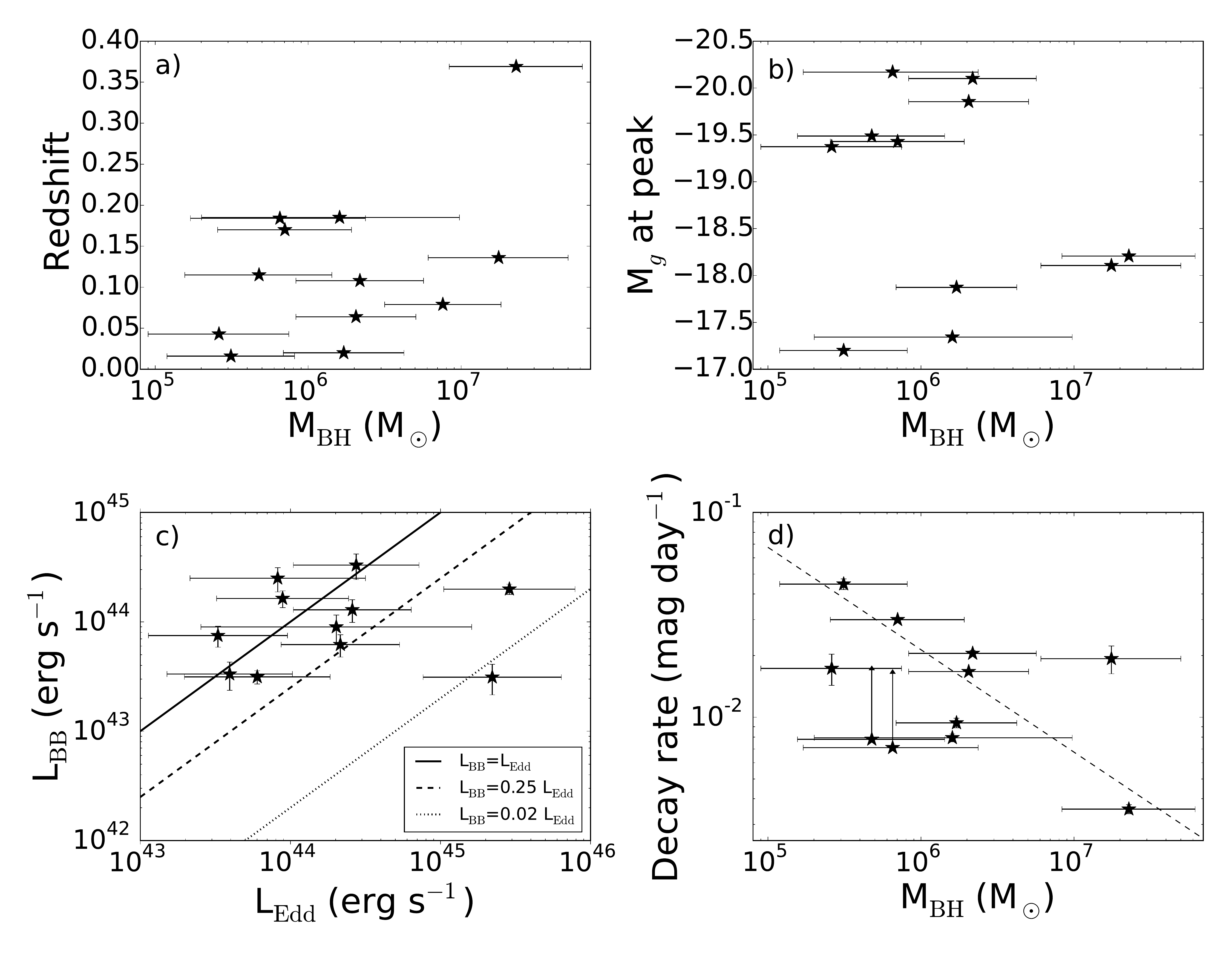}
  \caption{TDE observables as a function of black hole mass (or derived Eddington luminosity). Panel a) shows the host redshift as a function of M$_{\rm BH}$. Panel b) presents the (K-corrected) peak absolute magnitude as a function of M$_{\rm BH}$, while panel c) shows the peak blackbody luminosity as a function of the implied Eddington luminosity. The lines represent constant Eddington ratios. In panel d) we plot the decay rate (in the host rest frame) as a function of M$_{\rm BH}$. The dashed line represents the theoretically expected peak fallback rate (see text) and is proportional to M$_{\rm BH}^{-1/2}$.}
  \label{fig:correlations}
\end{figure*}

In panel b) of Fig. \ref{fig:correlations} we show the (K-corrected, \citeauthor{Humason1956} \citeyear{Humason1956}) peak absolute $g$-band magnitudes, i.e. the peak luminosity measured at 6.3\,$\times$\,10$^{14}$ Hz in the rest frame, plotted as a function of the black hole mass. We use the peak flux in the filter with the best temporal sampling in the literature, together with the blackbody temperature (taken from the literature, see Table \ref{tab:correlations}) to calculate the peak $g$-band magnitude in the rest frame of the host. Because we correct to the rest frame of the host galaxy, the specific filter choice is irrelevant. We note that for several TDEs we can only determine upper limits as the peak of the lightcurve was not observed. However, a visual comparison of the lightcurves of these events with other well sampled lightcurves of TDEs suggests that the peak was probably missed only by a few days and therefore the difference should be small. We do not observe a statistically significant trend of peak absolute magnitude with black hole mass. 

The observations suggest that current optical/UV surveys are already probing the fainter end of the TDE luminosity function (illustrated by the spread of optically/UV discovered TDEs between --17\,$\leq$\,M$_{\rm peak}$\,$\leq$\,--21) although it is likely that this luminosity function extends to even fainter sources. The bimodality in peak absolute magnitude is not significant and can be explained by small sample statistics. 

\subsubsection{Eddington ratio}
Using the blackbody temperature and the peak absolute magnitude, we calculate the integrated blackbody peak luminosity L$_{\rm BB}$. We determine the uncertainties by varying the temperature of the blackbody function within its errors. In panel c) of Fig. \ref{fig:correlations}, we compare L$_{\rm BB}$ to the Eddington luminosity implied by our black hole masses. The lines represent constant Eddington ratios, where the solid line represents the Eddington limit (i.e. where L$_{\rm BB}$\,=\,L$_{\rm Edd}$). The peak luminosity of all TDEs is consistent with being at the Eddington limit except for the two events with the highest black hole masses, which have Eddington ratios of $\sim$\,0.02 for TDE1 and 0.07 for D3--13. These properties are in agreement with simple dynamical predictions for the peak mass fallback rate $\dot{M}_{\rm peak}$, which give (e.g. \citealt{Stone2013})
\begin{equation}
\frac{\dot{M}_{\rm peak}}{\dot{M}_{\rm Edd}} \approx 130 \frac{\eta}{0.1} \left( \frac{M_{\rm BH}}{10^6M_\odot} \right)^{-3/2} \left( \frac{M_\star}{M_\odot} \right)^2 \left( \frac{R_\star}{R_\odot} \right)^{-3/2}
\end{equation}
Here $\eta$\,$\leq$\,1 is the radiative efficiency of the accretion flow produced by the tidal disruption of a star with mass M$_{\star}$ and radius R$_{\star}$ ($\dot{M}_{\rm Edd} \equiv L_{\rm Edd}\eta^{-1}c^{-2}$). In this scenario, the initial fallback rate is super-Eddington for low mass SMBHs and most stars on the main sequence. Nevertheless, if this simple fallback picture holds, the blackbody luminosity is limited to the Eddington luminosity. For a typical lower main sequence star (M$_{\star}$\,=\,0.3 M$_{\odot}$, R$_{\star}$\,=\,0.38 R$_{\odot}$), the initial fallback rate following disruption will be sub-Eddington when M$_{\rm BH}$\,$\geq$\,10$^{7.13}$\,M$_{\odot}$, as is probably the case for TDE1 and D3-13. In these flares, the fallback rate is likely sub-Eddington, and assuming that the luminosity tracks the fallback rate, so is the optical emission. If emission mechanisms other than blackbody operate, and depending on if these involve the emission of higher energy (e.g. X-ray) radiation, this picture could change drastically.

\subsubsection{Photometric evolution}
In Fig. \ref{fig:correlations} panel d), we plot the decay rate from the peak of the lightcurve as a function of M$_{\rm BH}$. Because of the heterogeneity of the available data, we use the best sampled lightcurve, which is either the {\it Swift} NUV filter or the optical $r$ or $g$ filters. The temperature evolution is observed to be near constant during the evolution of the flares \citep{Hung2017}. This means that the choice of filter should not impact these measurements significantly. The slope and its associated uncertainty are estimated using the standard formalism of linear regression. Although this may not be the model that best fits the data, it ensures that we can obtain a homogeneous set of measurements for all events. We also correct for the effect of time dilation in the observer's frame by scaling the measured decay rates with (1\,+\,$z$) to obtain the decay rates in the rest frame of the host galaxies \citep{Weinberg1972, Blondin2008}. 

The lowest mass black hole (iPTF--16fnl) hosted the fastest decaying TDE (see \citealt{Blago2017}), and the most massive black hole (D3--13) has the slowest decay timescale. The qualitative trend of a faster decay timescale with lower black hole mass as observed here is predicted by theory from the assumption that the peak optical luminosity traces only the peak mass fallback rate, which scales as $\dot{M}_{\rm peak} \propto M_{\rm BH}^{-1/2}$ \citep{Rees1988} and is plotted as a dashed line to guide the eye (note that this is not a fit to the data). However, the actual mechanism producing the optical emission is unknown and therefore it is unclear if a tight correlation should be expected. Other parameters such as the depth of the encounter (e.g. \citealt{Dai2015}), the properties of the star \citep{Lodato2009, Guillochon2013} or the spin of the black hole \citep{Kesden2012} may all influence the photometric evolution of the flare.

\subsection{The blackbody emission mechanism}
We use the blackbody temperatures and luminosities to estimate the blackbody radius where the emission is produced. If no uncertainty on the blackbody temperature is given in the literature, we assume it to be 10 per cent, similar to observed values (Table \ref{tab:correlations}). Uncertainties for the blackbody radius are obtained by standard error propagation, and do not include systematic errors. Because we have accurate constraints on the black hole masses, we investigate whether the estimated blackbody radii can discriminate between two current theoretical models for the optical emission. 

\begin{figure} 
  \includegraphics[width=0.5\textwidth, keepaspectratio]{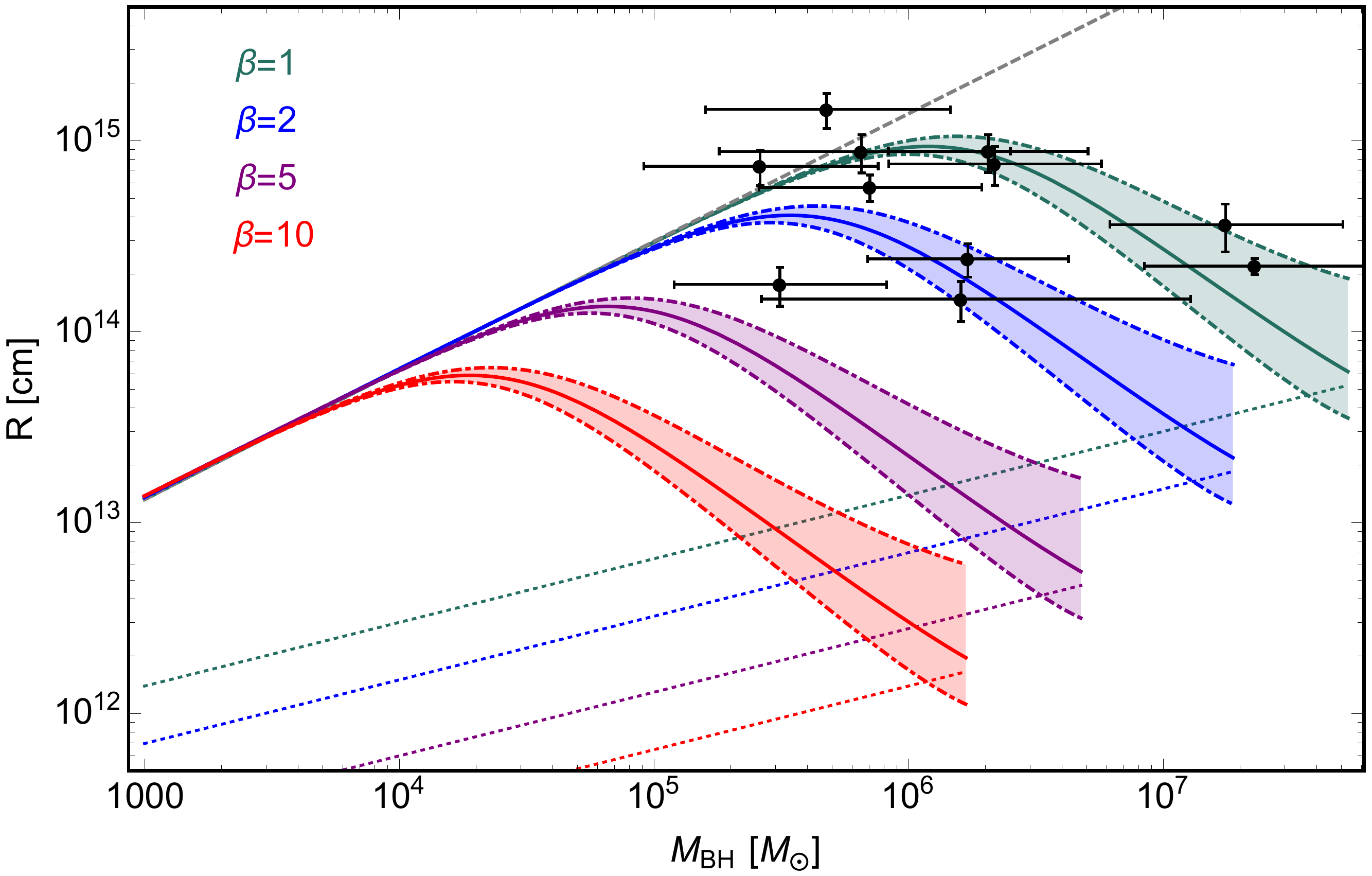}
  \caption{Blackbody radius as a function of M$_{\rm BH}$. Overplotted are different models for the origin of the blackbody emission for various $\beta$ values. The dotted lines represent a compact accretion disk at 2\,$\times$\,R$_{\rm p}$. The solid lines represent the stream self-intersection radius of a non-spinning black hole. All curves are for tidal disruptions of solar type stars. The shaded regions illustrates the effect of increasing black hole spin on the self-intersection radius. These regions are plotted out to a maximum $M_{\rm BH}$ corresponding to the Hills mass for a retrograde equatorial TDE around a BH with dimensionless spin parameter $a_{\rm BH}$\,=\,0.9. Somewhat larger SMBHs can still tidally disrupt solar type stars, but our post Newtonian predictions for the self-intersection radius would not be trustworthy for the most relativistic TDEs. The dashed grey line is the semi-major axis of the most tightly bound debris stream.}
  \label{fig:mbh_spin}
\end{figure} 
\begin{figure} 
\includegraphics[width=0.5\textwidth, keepaspectratio]{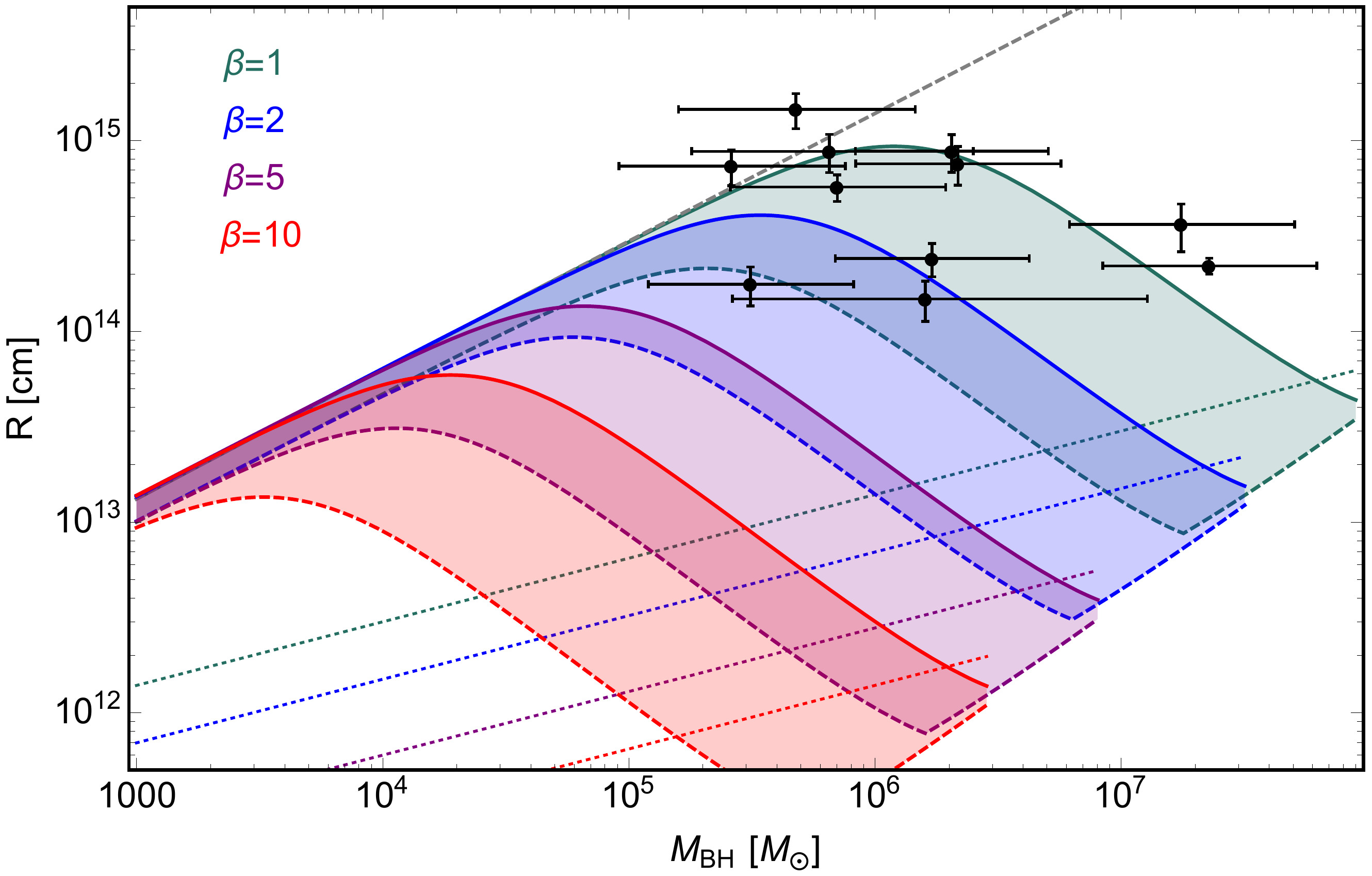}  
  \caption{Similar to Figure \ref{fig:mbh_spin}, but now illustrating the effect of varying stellar structure with the mass of the disrupted star on the stream self-intersection radius. We show the self-intersection radii of a 0.1 M$_{\odot}$ star (dashed coloured lines), a 1 M$_{\odot}$ star (solid lines) and everything in between.}
  \label{fig:mbh_mass}
\end{figure} 

We consider a model where the emission arises directly from a compact accretion disk, which forms at $\sim$\,2\,$\times$\,R$_{\rm p}$ (e.g. \citealt{Phinney1989}). 
Alternatively, we consider a class of models where the power source of the flare is dissipation of orbital energy in the circularization process \citep{Lodato2012}, and the blackbody emission originates in shocks at the stream self-intersection radius \citep{Piran2015}. Stream self-intersection is caused by general relativistic apsidal precession, and scales steeply with the ratio of R$_{\rm p}$ to the gravitational radius R$_{\rm g}=GM_{\rm BH}/c^2$. For this reason, \citet{Dai2015} argue that shallow encounters (at low $\beta = R_{\rm T} / R_{\rm p}$, the penetration factor of the fatal orbit) circularize relatively far from the BH, leading to optical/UV emission, while high $\beta$ encounters produce X-ray TDEs.

We estimate the self-intersection radius R$_{\rm SI}$ by considering the orbits of test particles around a SMBH. Averaged over one orbit, general relativistic apsidal precession causes the argument of pericenter $\omega$ to advance by an amount
\begin{equation}
\delta \omega = A_{\rm S}-2A_{\rm J} \cos \iota,
\end{equation}
at leading post-Newtonian order. In this equation, the contributions to apsidal precession from the black hole mass and spin-induced frame dragging are A$_{\rm S}$ and A$_{\rm J}$, respectively, and are given by \citep{Merritt2010}
\begin{align}
A_{\rm S} & =\frac{6\pi}{c^2}\frac{GM_{\rm BH}}{R_{\rm p}(1+e)} \approx 11.5^{\circ} \left( \frac{\tilde{R}_{\rm p}}{47.1} \right)^{-1} \\
A_{\rm J} & =\frac{4\pi a_{\rm BH}}{c^3}\left( \frac{GM_{\rm BH}}{R_{\rm p}(1+e)} \right)^{3/2}  \approx 0.788^{\circ} \left( \frac{\tilde{R}_{\rm p}}{47.1} \right)^{-3/2} a_{\rm BH}
\end{align}
In the above equations, the orbital pericenter, eccentricity, and inclination (with respect to the BH equatorial plane) are R$_{\rm p}, e,$ and $\iota$, respectively. The BH possesses a mass M$_{\rm BH}$ and a spin $a_{\rm BH}$. Likewise, $\tilde{R}_{\rm p}$ is the orbital pericenter normalized by the gravitational radius R$_{\rm g}$, and $\tilde{R}_{\rm p}= 47.1$ for a $10^6$ \,M$_{\odot}$ SMBH. The approximate equalities on the right assume highly eccentric orbits ($1+e \approx 2$).

We now limit ourselves to the case of coplanar orbits, i.e. we assume that the orbital plane of the star is perpendicular to the spin axis of the BH. If we assume apsidal precession occurs impulsively at pericenter, we find that the debris stream will self intersect at a distance \citep{Dai2015}
\begin{equation}
R_{\rm SI}=\frac{R_{\rm p}(1+e)}{1+e\cos(\pi + \delta \omega/2 )} \label{eq:rSI}
\end{equation}
Stream self-intersection may be greatly complicated by inclined orbits undergoing nodal precession \citep{Guillochon2015, Hayasaki2016}, but this is primarily due to small vertical offsets between debris streams; the projected radius of self-intersection will not deviate greatly from Eq. \ref{eq:rSI} unless $\tilde{R}_{\rm p} \sim 1$. In computing the depth $\beta$ of each encounter, we take the tidal radius R$_{\rm T} \equiv\ $R$_\star $(M$_{\rm BH}$/M$_\star)^{1/3}$. Here M$_\star$ and R$_\star$ are the mass and radius of the victim star, respectively, and we assume the lower main sequence relationship R$_\star \propto\ $M$_\star^{0.8}$ \citep{Kippenhahn1990}.

In Figure \ref{fig:mbh_spin} we show the expected emission region in the case of the compact accretion disk model (dotted lines), while the solid (dot-dashed) lines represent Schwarzschild (Kerr) stream self-intersection radii. The shaded areas illustrate the effect of increasing black hole spin ($a_{\rm BH}$), while the different colours represent different impact parameters, with $\beta$\,$\approx$\,1 being the most common type of event \citep{Stone2016}. The shaded areas below the solid lines represent retrograde spin values ($a_{\rm BH}$\,$\leq$\,0), while the area above the solid line corresponds to prograde spins ($a_{\rm BH}$\,$\geq$\,0). A retrograde spin increases the amount of apsidal precession, which decreases the stream self-intersection radius \citep{Dai2015}. Conversely, a prograde spin diminishes the apsidal precession, forcing a self-intersection at larger radius. Our mass and radius measurements are overplotted as black dots.

The dotted lines in Figure \ref{fig:mbh_mass} are the same as in Figure \ref{fig:mbh_spin}, while the dashed and solid lines illustrate the effect of stellar mass; here the mass of the disrupted star is M$_{\star}$\,=\,0.1 M$_{\odot}$ and M$_{\star}$\,=\,1 M$_{\odot}$, respectively. In this case we have assumed a non-spinning (Schwarzschild) BH. 

Our inferred blackbody radii, which can be interpreted as the location from which the blackbody emission (at peak brightness) originates, are consistent with the self-intersection radius of shallow impact encounters ($\beta$\,$\sim$\,1--2), regardless of the BH spin or mass of the disrupted star. A scenario involving an accretion disk which extends to a few tens of gravitational radii from the black hole can be ruled out as the origin of the blackbody luminosity at peak brightness by our measurements. It is clear from Figures \ref{fig:mbh_spin} and \ref{fig:mbh_mass} that the stream self-intersection radius (at fixed M$_{\rm BH}$) is more sensitive to the mass of the disrupted star than it is to increasing black hole spin. While the degeneracy between $a_{\rm BH}$ and M$_{\star}$ precludes us from inferring the specific combination of black hole spin, impact parameter and stellar mass of the TDEs in our sample, it does allow us to conclude that the most likely region of origin for the blackbody emission for {\it all} optical/UV TDEs is at the stream self-intersection radius of low $\beta$ encounters, lending empirical support to the stream-stream collision model for the power source of optical TDEs at peak brightness. However, we note that -- while this data is deeply inconsistent with simple models of compact accretion disks -- accretion-powered reprocessing models may still be able to explain the observed optical photospheres {\it provided that the reprocessing layer is formed near the stream self-intersection point.} The circularization process is still poorly understood, but our results suggest that accretion-powered reprocessing models will only remain viable explanations for TDE optical emission if debris circularization naturally produces optically thick photospheres on self-intersection scales.

The shock-powered model of \citet{Piran2015} predicts that for a circularization powered flare the peak luminosity should depend only weakly on M$_{\rm BH}$, in agreement with our observations (panel b, Fig. \ref{fig:correlations}). This model also naturally explains the shrinking of the observed blackbody radius over time \citep{Hung2017} as an inward drift of the shock after debris that has passed through pericenter settles into more circular orbits \citep{Piran2015}. However, we do not find a clear correlation between the blackbody temperature and black hole mass as predicted by the same model.

It is important to keep in mind that the precise value of the stream self-intersection radius depends on the combination of parameters $\beta$, $a_{\rm BH}$ and mass of the disrupted star. We note that a complete disruption requires $\beta$\,$\gtrsim$\,1.85 for low mass stars, and $\beta$\,$\gtrsim$\,0.95 for Sun-like stars \citep{Guillochon2013}.
Although all the sources in Figure \ref{fig:mbh_mass} are consistent with this criterion, the figure suggests that some TDEs are due to low $\beta$ encounters of stars near the high mass end of the stellar mass function (M$_{\star}$\,$\approx$\,1 M$_{\odot}$) rather than due to 0.3 M$_{\odot}$ stars, as expected from the initial mass function \citep{Kochanek2016}. It is unclear if a selection bias in the current TDE sample could cause this tension. On the other hand, we remark that a non-zero, prograde BH spin can increase the self-intersection radius at given $\beta$ and disrupted stellar mass. We speculate that the discrepancy could decrease if some of the SMBHs in our sample have non-zero prograde spins. 

\subsection{Implications for the TDE rate}
\label{sec:rates}
Based on theoretical arguments, it has been proposed that the rate of TDEs should be dominated by the lowest mass galaxies hosting black holes \citep{Magorrian1999, Wang2004, Stone2016}. It is unclear how this theoretical TDE rate translates into a \textit{observed} TDE rate. At present, there is a strong tension between the observed ($\sim$\,10$^{-5}$\,Mpc$^{-3}$ yr$^{-1}$, e.g. \citealt{Donley2002}, \citealt{vanVelzen14}, \citealt{Holoien2016}) and theoretical ($\sim$\,10$^{-4}$\,Mpc$^{-3}$ yr$^{-1}$, e.g. \citealt{Magorrian1999}, \citealt{Wang2004}) TDE rates. \citet{Stone2016} study the effect of a number of parameters and assumptions that go into the theoretical and observational rate calculations, and conclude that there is no straightforward way to bring the two closer together. 

Our mass distribution (Figure \ref{fig:massdistrib}) shows that the observations qualitatively agree with the theoretical expectation that the sample of optical TDEs should be dominated by disruptions in galaxies hosting low mass ($\sim$\,10$^6$ M$_{\odot}$) black holes (see e.g. fig. 6 in \citeauthor{Kochanek2016} \citeyear{Kochanek2016}). The fact that we observe TDEs in lower mass black holes than previously assumed has important consequences for the inferred TDE rate. In particular, there are a number of physical mechanisms that can act to reduce the TDE luminosity (and thus observed rate) for BH masses below $\sim$10$^{6.5}$ M$_{\odot}$. For example, \citet{Guillochon2015} argue that inefficient circularization affects the TDE energy output for M$_{\rm BH}$\,$\leq$\,10$^6$\,M$_{\odot}$, while \citet{Metzger2016} suggest that adiabatic losses in a slow and dense outflow may reduce the blackbody luminosity of TDEs around 10$^6$\,M$_{\odot}$ black holes. However, our work illustrates that the current TDE sample is dominated by $\sim$\,10$^6$\,M$_{\odot}$ black holes and contains several BHs with lower masses. Therefore, the current rate estimates apply to this low mass regime and cannot be invoked to explain the discrepant TDE rates. In other words, we find the possibility of a hidden population of TDEs around low mass (10$^{5-6}$ M$_{\odot}$) BHs as an explanation for the rate discrepancy unlikely. Moreover, his is further supported by the fact that we do not observe a strong correlation between the TDE peak luminosity and black hole mass, which implies that any selection effect due to the low volume probed by TDEs around low mass BHs does not significantly affect the current sample (at least down to M$_{\rm BH}$\,$\sim$\,10$^6$\,M$_{\odot}$). 

\subsection{Intermediate mass black holes}
\label{sec:imbh}
Our TDE selected host galaxy sample suggests that there is a large, hidden population of low mass black holes lying dormant in the centers of galaxies. Low mass black holes are notoriously hard to find, even when they accrete from a steady reservoir of gas. Some searches exploit the short timescales of X-ray variability to separate low from high mass black holes (e.g. \citeauthor{Greene2007} \citeyear{Greene2007}). Alternatively, scaling relations based on optical emission lines \citep{Kauffmann2003} or virial based techniques can be used to estimate M$_{\rm BH}$ in active galaxies \citep{Reines2013}. \citet{Kauffmann2003} show that the AGN fraction in low mass galaxies in the local Universe (0.02\,$\leq$\,$z$\,$\leq$\,0.3) does not rise above a few per cent, while \citet{Gallo2010} find that the AGN fraction decreases with increasing SMBH mass. The large majority ($\geq\,95\,\%$) of black holes in low mass galaxies are therefore currently hidden from our view, and TDEs can be a powerful tool to find and study the demographics of low mass galaxies and their low mass central SMBHs.

If the mass distribution of our sample of TDE hosts is representative for the population of all optical/UV TDE host galaxies, this holds exciting prospects for finding intermediate mass black holes in the local universe. In the near future, optical surveys such as performed by the Zwicky Transient Factory (ZTF), Gaia and the Large Sky Synoptic Telescope (LSST) are expected to uncover thousands of TDEs and thus large numbers of low mass black holes. This can open up a new avenue for the systematic study of IMBH formation and evolution, and the galaxies in which they reside. Using TDEs as an independent probe for BHs in low mass galaxies, mass measurements on this future sample of TDE host BHs will shed light on the validity of the M\,--\,$\sigma$ relation at the low end (see Figure \ref{fig:msigma_overplotted}), and will help constrain the black hole occupation fraction at the low mass end. The existence and masses of IMBHs in low mass galaxies are an important tool to differentiate between SMBH formation scenarios (e.g. \citeauthor{VolonteriLodato2008} \citeyear{VolonteriLodato2008}), and can enable the study of the main mechanisms for low mass SMBH growth and evolution as well as their formation. For example, different seed models leave different (and observable) imprints on the current ($z$\,=\,0) MBH mass function \citep{Volonteri2010}. 

\section{Conclusions}
\label{sec:summary}
We present the first systematic black hole mass measurements for a sample of TDE host galaxies in the Northern sky using the M\,--\,$\sigma$ relation. Our host galaxy sample of optically/UV selected TDEs encompasses 12 sources, and is complete down to g$_{\rm host}$\,=\,22 mag, spanning a redshift range between 0.016 and 0.37. We use medium resolution spectroscopic observations in combination with the penalized pixel fitting routine to extract the line of sight velocity distributions, and in particular the velocity dispersions. Care is taken to correct for the instrumental broadening, and we study the effect of using the luminosity-weighted LOSVD as a proxy for the central velocity dispersion. We find that the luminosity-weighted LOSVD agrees well with the central velocity dispersions.

Using the M\,--\,$\sigma$ relation from \citet{Ferrarese2005} we convert the velocity dispersion measurements into black hole masses. Our galaxies host black holes with masses ranging between 3$\times$10$^5$\,M$_{\odot}$ $\leq$ M$_{\rm BH}$ $\leq$ 2$\times$10$^7$\,M$_{\odot}$. Our mass distribution agrees with theoretical estimates; the optical TDE population is dominated by low mass ($\sim$\,10$^6$ M$_{\odot}$) black holes. We find suggestive evidence for a correlation between the black hole mass and redshift, which is expected for a flux-limited sample. Furthermore our observations reveal tentative evidence for a correlation between the photometric evolution timescale (decay rate) and the mass of black hole: TDEs around lower mass black holes evolve faster. We note that these correlations are not statistically significant, potentially due to both the uncertainties on the observables and the small sample size. The blackbody emission of our sources is consistent with being at the Eddington limit at peak brightness, except for the two sources with M$_{\rm BH}$\,$\geq$\,10$^{7.1}$\,M$_{\odot}$ for which the Eddington ratio is $\leq$\,0.1. These properties corroborate the standard TDE picture as a satisfactory explanation for these events.

Regarding the origin of the blackbody emission, we compare the blackbody radii of the flares with models proposed to explain the origin of the emission, including a compact accretion disk and shocks due to stream self-intersections. We find that the emission region at peak brightness is located more than $\sim$100\,R$_{\rm g}$ from the black holes, and is consistent with the stream self-intersection radius of disruptions at low $\beta$\,$\sim$\,1\,--\,2. This rules out a compact accretion disk as the direct origin of the blackbody emission, and suggests that at peak luminosity, TDEs are powered by shocks due to stream-stream collisions rather than directly by accretion power.

Finally, our finding that TDEs frequently occur in low mass ($\sim$\,10$^{6}$\,M$_{\odot}$) black holes implies a worsening of the rate discrepancy between theoretical and observational rates. This follows by noting that several mechanisms predict a lower flare brightness for TDEs in low mass $\leq$\,10$^{6.5}$\,M$_{\odot}$ BHs, while our observations show that the current TDE sample is dominated by such events. This may not be true if the currently observed TDE rate is only a small fraction of the true TDE rate (e.g. due to other selection effects).
 
Our results suggest that there is a large population of dormant, low mass black holes hidden at the centres of local galaxies. TDEs could provide an opportunity to uncover this population through (near-) future time domain surveys, which are expected to find thousands of TDEs per year. The sample of TDE host galaxies may be useful to constrain the properties of low mass black holes, as well as the formation channels and dominant growth and feeding mechanisms of SMBHs.

\section*{Acknowledgements}
We would like to thank the anonymous referee for suggestions that improved the manuscript.
TW wishes to thank D. Lena for assisting in the observing runs, and is grateful to D. Lena and M. Torres for useful discussions. SvV is supported by NASA through a Hubble Fellowship (HSTHF2-51350). PGJ acknowledges support from European Research Council Consolidator Grant 647208. NCS acknowledges support through NASA from Einstein Postdoctoral Fellowship Award Number PF5-160145. The research leading to these results has received funding from the European Union’s Horizon 2020 Programme under the AHEAD project (grant agreement n. 654215). SG is supported in part by NSF CAREER grant 1454816 and NASA Keck Grant 1568615. Part of this work was inspired by discussions within International Team \#371 Using Tidal Disruption Events to Study Super-Massive Black Holes at the International Space Science Institute in Bern, Switzerland. We thank Tom Marsh for developing the software package \textsc{molly}. The WHT is operated on the island of La Palma by the Isaac Newton Group of Telescopes in the Spanish Observatorio del Roque de los Muchachos of the Instituto de Astrofisica de Canarias. The ISIS spectroscopy was obtained as part of programmes W15BN010 and W16AN007. Based on observations made with ESO Telescopes at the La Silla Paranal Observatory with the director's discretionary time, programme ID 297.B-5062(A) (P.I. Jonker). Some of the data presented herein were obtained at the W.M. Keck Observatory, which is operated as a scientific partnership among the California Institute of Technology, the University of California and the National Aeronautics and Space Administration. The Observatory was made possible by the generous financial support of the W.M. Keck Foundation. We wish to recognize and acknowledge the very significant cultural role and reverence that the summit of Mauna Kea has always had within the indigenous Hawaiian community. We are most fortunate to have the opportunity to conduct observations from this mountain.

\bibliographystyle{mnras.bst}
\bibliography{bibliography_tdehosts.bib}
\label{lastpage}
\end{document}